\documentclass[pra,aps,
superscriptaddress,
reprint,
footinbib,
notitlepage,
twocolumn,
noamsmath,
longbibliography
]{revtex4-2} 
\usepackage{pgfkeys}
\usepackage{mathtools}
\usepackage{txfonts}
\usepackage{soul}
\usepackage{graphicx}
\usepackage{dcolumn}
\usepackage{bm}
\usepackage{bbm}
\usepackage{hyperref}
\usepackage[mathlines]{lineno}

\usepackage{url}
\usepackage[dvipsnames]{xcolor}
\usepackage{slashed}

\usepackage{amsfonts}
\usepackage{graphicx}
\usepackage{hyperref}
\usepackage{comment}

\hypersetup{
	colorlinks   = true, 
	urlcolor     = blue, 
	linkcolor    = blue, 
	citecolor   = red 
}

\newcommand{\eqnref}[1]{Eq.~(\ref{#1})}
\newcommand{\figref}[1]{Fig.~\ref{#1}}
\newcommand{\sfigref}[2]{Fig.~\hyperref[#1]{\ref{#1}#2}}
\newcommand{\tabref}[1]{Table~\ref{#1}}
\newcommand{\secref}[1]{Sec.~\ref{#1}}

\newcommand{\abs}[1]{\left| #1 \right|}

\newcommand{\beq}{\begin{equation}}
	\newcommand{\eeq}{\end{equation}}
\newcommand{\beqd}{\begin{equation*}}
	\newcommand{\eeqd}{\end{equation*}}
\newcommand{\dou}{\partial}

\newcommand{\expect}[1]{\left\langle #1 \right\rangle}
\newcommand{\bra}[1]{{{\langle #1 |}}}
\newcommand{\ket}[1]{{| #1 \rangle}}
\newcommand{\bpm}{\begin{pmatrix}}
	\newcommand{\epm}{\end{pmatrix}}

\newcommand{\D}[1]{\textup{d}{#1}} 
\newcommand{\tr}{\textup{tr}}

\newcommand{\bcen}{\begin{center}}
	\newcommand{\ecen}{\end{center}}
\newcommand{\btab}{\begin{tabular}}
	\newcommand{\etab}{\end{tabular}}
\newcommand{\bdes}{\begin{description}}
	\newcommand{\edes}{\end{description}}

\newcommand{\bea}{\begin{eqnarray}}
	\newcommand{\eea}{\end{eqnarray}}

\newcommand{\half}{\frac{1}{2}}
\newcommand{\bary}{\begin{array}}
	\newcommand{\eary}{\end{array}}
\newcommand{\benum}{\begin{enumerate}}
	\newcommand{\eenum}{\end{enumerate}}
\newcommand{\bitem}{\begin{itemize}}
	\newcommand{\eitem}{\end{itemize}}

\newcommand{\bv} { \mbox{\boldmath $v$}}

\newcommand{\bx} { {\boldsymbol{x}}}

\newcommand{\bD}{{\boldsymbol{D}}}

\newcommand{\mean}[1]{\langle #1 \rangle}

\newcommand{\SMRef}[1]{{see \cite{SM}, \protect{#1}}}

\newcommand{\Integers}{{\mathbb{Z}}}
\newcommand{\ci}{\mathbbm{i}}

\def\makeSM{1}
\ifdefined\makeSM
\else
\include{fcsedge_SM.aux}
\fi

\newcommand{\paraheading}[1]{\noindent{\em #1:}}

\newcommand{\mytitle}{Fractons on the edge}
\newcommand{\authorZero}{Bhandaru Phani Parasar}
\newcommand{\authorOne}{Yuval Gefen}
\newcommand{\authorTwo}{Vijay B.~Shenoy}

\begin{document}
	
	\title{Fractons on the edge}

	\author{\authorZero}
	\email{bhandarup@iisc.ac.in}
	\affiliation{Centre for Condensed Matter Theory, Department of Physics, Indian Institute of Science, Bangalore 560012, India}
	
	\author{\authorOne}
	\email{yuval.gefen@weizmann.ac.il}
	\affiliation{Weizmann Institute of Science, 234 Herzl St.,
PO Box 26,
Rehovot 7610001, 
Israel}
  
      \author{\authorTwo}
      \email{shenoy@iisc.ac.in}
	\affiliation{Centre for Condensed Matter Theory, Department of Physics, Indian Institute of Science, Bangalore 560012, India}

	\begin{abstract}
We develop a theory of edge excitations of fractonic systems in two dimensions, and elucidate their connections to bulk transport properties and quantum statistics of bulk excitations.
The system we consider has immobile point charges, dipoles constrained to move only along lines perpendicular to their moment, and freely mobile quadrupoles and higher multipoles, realizing a bulk fractonic analog of fractional quantum Hall phases. 
We demonstrate that a quantized braiding phase between two bulk excitations is obtained only in two cases: when a point quadrupole braids around an immobile point charge, or when two non-orthogonal point dipoles braid with one another. The presence of a boundary edge in the system entails {\em two} types of gapless edge excitation modes, one that is fractonic with immobile charges and longitudinal dipoles, and a second non-fractonic mode consisting of transverse dipoles. We derive a novel current algebra of the fractonic edge modes. 
Further, investigating the effect of local edge-to-edge tunneling on these modes, we find that such a process is a relevant perturbation suggesting
the possibility of edge deformation. 
	\end{abstract}
    
	\maketitle

\paraheading{Introduction} Following the seminal works\cite{Wen1990,Kitaev2009,Ryu2010} on classification of quantum phases of matter based on topology and entanglement\cite{Hasan2010,Qi2011,Chiu2016,Wen2017}, the discovery of fracton phases\cite{Chamon2005,Castelnovo2012,Haah2011,Vijay2015,Vijay2016} has kindled wide-ranging interest\cite{Nandkishore2019,PretkoChenYou2020}. Dubbed fractons, the excitations of these phases are immobile or have ``fractional'' (i.~e., restricted to a line/plane) mobility. These phases offer new platforms for the realization of stable quantum memories\cite{Haah2011,Nandkishore2019,PretkoChenYou2020} even as they continue to provide intriguing theoretical questions and challenges.

First formulated using discrete lattice models\cite{Chamon2005,Castelnovo2012,Haah2011,Bravyi2013,Vijay2015,Vijay2016}, fractonic systems host a rich and remarkable variety of phenomena, in contrast with the conventional topologically ordered systems such as the toric code\cite{Kitaev2003}.  For example, in addition to the mobility restriction of excitations, they possess other notable features such as sub-extensive ground state degeneracy\cite{Vijay2015}, strong dependence on the microscopic details of the models such as the type of lattice\cite{Slagle2018,Shirley2018}, etc. The physics behind the quantum statistics of excitations is also substantially richer\cite{PaiHermele2019,SongNathananShirleyHermele2024} owing to the mobility restrictions.
The continuum description of these phases requires a tensor gauge theory description
\cite{Xu2006, RYX2016, Pretko2017Mar, Pretko2017Jul, Pretko2017Sep,SlagleYBKim2017, YouDevakulSondhiBurnell2020, ShenoyMoessner2020,
Slagle2021,GorantlaLamSeibergShao_PRB2023,Spieler2023}. 
Intriguing connections between such gauge theories and other systems are known via duality theory\cite{Zaanen2004, Beekman2017, PretkoRadzihovsky2018, PretkoRadzihovsky2018a, Gromov_PRL2019, Gromov2019, SeibergShao2021, Manoj2021, DoshiGromov2021, NguyenMoroz2024arXiv}. General classification of these phases\cite{Gromov_PRX2019,Radicevic2019} has also been explored.

 Notwithstanding the advances noted above 
 there are key outstanding questions in the context of fractons.  Do the edges (or surfaces in higher dimensions) of fractonic systems host gapless excitations, as found, for example, in two-dimensional topological systems (e.~g.~quantum Hall phases\cite{Wen_IJMPB1992}, topological insulators\cite{Hasan2010,Qi2011})? Provided the answer is in the affirmative, how do the properties of the edge excitations reflect the fractonic properties of the bulk, analogous to bulk-edge relations in topological phases.  
Furthermore, what is the fate of the edge excitation in the presence of imperfections that cause edge-to-edge local tunneling?

This letter addresses these questions  in a continuum field theoretical framework 
focusing on the role of the edge in manifesting the universal features of two-dimensional fractonic systems. The latter include braiding statistics, and transport properties.  
Our key findings are: 
\begin{enumerate}
\item  Noting that the bulk excitations in our theory comprise fully immobile charges, dipoles that can move only transverse to the direction of their dipole moment, and fully mobile quadrupoles (and higher multiples),  a fractional statistical phase is obtained only in two cases:  first when a mobile point quadrupole braids around an immobile point charge, and second when two non-orthogonal dipoles perform a braiding spacetime trajectory. 
\item The fractonic Hall-like response of the system is related to the statistical phase of bulk excitations, similar to what is found in fractional quantum Hall systems.
\item  Quite remarkably, in a system with edges, we find that there are {\em two gapless edge modes}. The first edge mode corresponds to a one-dimensional fractonic system with conserved charges, dipoles, and quadrupoles. The second gapless mode has non-fractonic gapless degrees of freedom. We characterize these excitations by their quantum commutation relations that turn out to be related to the statistical phase of bulk excitations. 
\item We find that certain edge excitations are susceptible to edge-to-edge tunneling, which turns out to be a relevant perturbation. This suggests a possibility of edge deformation similar to that found in fractional Hall effect\cite{MoonKaneGirvinFischer1993,Fendly1995}.

\end{enumerate}
This work reveals the intricate connection between the statistics of bulk excitations, transport properties, and the nature of the gapless edges of fractonic systems, providing a clarifying theoretical picture.
Technically, this is achieved by a field-theoretical approach that employs a fractonic Chern-Simons (FCS) model\cite{Pretko2017c,PremPretkoNandkishore,Fliss2021}, with rank two tensor gauge fields and is parameterized by an integer level $k$, in a two-dimensional spatial setting as detailed below.

\begin{figure}
    \centering
    \includegraphics[width=0.99\linewidth]{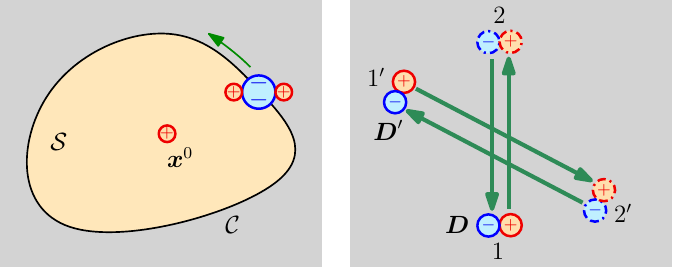}
    \centerline{(a)~~~~~~~~~~~~~~~~~~~~~~~~~~~~~~~~~~~~~~~~~~~~~~~~~~~(b)}
    \caption{{\bf Braiding Statistics:} (a) Braiding of a point quadrupole with $\tr{\Theta_{ij}} = \Theta \ell$ around an immobile point charge $Q \ell^{-1}$ located at $\bx^0$. The quadrupole traces a closed path ${\cal C}$, which encloses the area ${\cal S}$. (b) Braiding of dipoles $\bD$ and $\bD'$ with ``crossing paths'' with the following sequence: $\bD$ from $1$ to $2$, $\bD'$ from $1'$ to $2'$, $\bD$ from $2$ to $1$, $\bD'$ from $2'$ to $1'$.  The path $1 \to 2$ is orthogonal to $\bD$, and $1' \to 2'$ is orthogonal to $\bD'$. }
    \label{fig:stat_phase}
\end{figure}

\paraheading{Fractonic Chern-Simons Theory} We adopt a continuum gauge theoretical description\cite{Pretko2017c,PremPretkoNandkishore,Fliss2021} based on gauge fields $(\phi, a_{ij})$, where $a_{ij}$ is a traceless 2-nd rank symmetric tensor, in 2 spatial (coordinates $x_i$, $i=1,2$) and 1 time ($t$) dimensions. The electric $e_{ij}$ and magnetic $b$ fields associated with these gauge fields 
\beq\label{eqn:EB}
\begin{split}
e_{ij} & = \dou_i \dou_j \phi - \half \delta_{ij} \dou_l \dou_l \phi - \dou_t a_{ij}, \;\;\;
b = \epsilon_{jl} \dou_j \dou_i a_{li}
\end{split}
\eeq
where $\delta_{ij}$ and $\epsilon_{jl}$ are Kronecker delta and alternating tensor respectively; Einstein summation convention is adopted. These fields are invariant under the  U$(1)$ gauge transformation 
\beq\label{eqn:GT}
\phi \to \phi + \dou_t \xi, \;\;a_{ij} \to a_{ij} + \left(\dou_{i} \dou_j \xi  - \half \delta_{ij }\dou_l \dou_l \xi\right)
\eeq
where $\xi(t,x_i)$ is an arbitrary  space-time function. The action dubbed fractonic Chern-Simons (FCS) action, an effective theory for the Hall effect of dipoles\cite{PremPretkoNandkishore}, is
\beq\label{eqn:FCSaction}
S_{\textup{FCS}}[\phi,a_{ij}]= \frac{k}{4\pi} \int  \D{t}\D{^2 x} \left( b \phi  - \epsilon_{ij} e_{jl} a_{li}\right) +  \int  \D{t} \D{^2 x} \left(\rho \phi - J_{ij} a_{ij} \right)
\eeq
where $\rho$ and symmetric trace-free $J_{ij}$  are the charge and current densities respectively. The theory is not a topological field theory, compared to  the usual U$(1)$ Chern-Simons theory.  The Bianchi identity
$$
\dou_t b + \epsilon_{jl} \dou_j \dou_i e_{li} = 0
$$
and the continuity equation 
\beq\label{eqn:Continuity}
\dou_t \rho + \dou_i \dou_j J_{ij} = 0
\eeq
ensures that the theory is invariant under the gauge transformation \eqnref{eqn:GT}. Further, based on invariance under large gauge transformations,  it can be shown that $k$ is an integer (\SMRef{\secref{SM:sec:quantk}}). The continuity equation implies that the charge, dipole moment, and trace of the quadrupole moment are conserved. This imposes mobility restrictions: point charges are fully immobile, and point dipoles are mobile only in the direction transverse to their dipole moment (i.~e., they are lineons \cite{Shirley2019a}).
In the convention adopted here, $[\rho]=\textup{L}^{-3}$, i.~e., the strength of an elementary point charge is $Q \ell^{-1}$,   a point dipole is $D$  (dimensionless), and a point quadrupole tensor $\Theta_{ij}$ has trace $ \Theta_{ii} = \Theta \ell$, where $Q$, $D$ and $\Theta$ are integers and $\ell$ is a microscopic length scale such as the lattice parameter of the underlying ultraviolet theory(\SMRef{\secref{SM:sec:chargeQuantization}}).

\paraheading{Statistics of Bulk Excitations and Bulk Response} We show here that there are only two types of processes of braiding point excitations around each other, consistent with the mobility restrictions, that produce a nontrivial statistical phase. In the first process, the a quadrupole  $\Theta_{ij}$,  is braided around a static fracton charge $Q \ell^{-1}$. Using the current $J_{ij}$ associated with the motion of a quadrupole (\SMRef{\secref{SM:sec:quadrupoleLagrange}}), we obtain the phase picked up by the quadrupole to be $\exp\left(\ci \frac{\ell \Theta}{2} \int_{{\cal S}} \D{^2x} \,  b \right)$ where the integral is over the area ${\cal S}$ enclosed by the loop ${\cal C}$ shown in \figref{fig:stat_phase}(a). For an isolated fracton charge of strength $Q$ located at $x^0_i$ inside the loop, $b(x) =- \frac{2\pi}{k} \delta(x_i - x^0_i)$, this results in a statistical phase $\exp\left(-\ci \frac{\pi}{k} Q \Theta\right)$. Since $Q$ and $\Theta$ are integers for elementary charges and quadrupoles, the statistical phase is a rational multiple of $2\pi$. Interestingly, no other mobile higher multipole picks up a statistical phase when taken around a fracton charge. Further, if the point charge at $x^0_i$ is replaced by any point higher multipole, no statistical phase is picked up. In the second process, two dipoles follow close loop trajectories that effectively produce space-time braiding, as shown in \ref{fig:stat_phase}(b). For two dipoles with moments $\bD$ and $\bD'$ with ``crossing paths'' as shown in \figref{fig:stat_phase}, we find  that the statistical phase is $\exp({\ci \frac{2\pi}{k} \bD \cdot \bD'})$ (\SMRef{\secref{SM:sec:dipoleStatPhase}}). Since $\bD \cdot \bD'$ is an integer, the statistical phase is again a rational fraction of $2 \pi$, determined by $k$.
The two cases found here are reminiscent of observations in a non-fractonic model with short-time fractonic behavior of ref.~\cite{DelfinoFontanaGomesChamon2023}. The properties found are to be contrasted with the usual Chern-Simons theory (e.g.~of fractional Hall effect, \cite{Wen_IJMPB1992}) where only charges pick up phases when braided around each other. On the other hand, the physics of FCS is similar to the usual CS in that the statistical phase, when present, is determined by the level $k$.

Preparing the ground to study the bulk responses, and the analysis of edge modes, we recast the theory using {\em background gauge fields} $(\Phi,A_{ij})$ (analogous to the Maxwell field in the standard Chern-Simons theory). The gauge fields couple to the ultraviolet fractonic charges and currents -- ``ultraviolet matter'' -- $(\rho^u,J^u_{ij})$. 
The latter can be expressed using the dynamical gauge fields  $(\phi,a_{ij})$ as
\beq
\rho^u = \frac{1}{2 \pi} b, \;\;\; J^u_{ij} = \frac{1}{2\pi} \frac{\epsilon_{il} e_{lj}+\epsilon_{jl} e_{li}}{2}.
\eeq
so that continuity equation $\dou_t \rho^u_{ij} + \dou_i \dou_j J^u_{ij} = 0$ is identically satisfied. 

The theory with the ultraviolet matter coupled to the background gauge fields can now  be written as
\begin{align}
\nonumber
    S_{\textup{FCS-BG}}[\phi, a_{ij};\, & \Phi, A_{ij}]=\frac{k}{4\pi} \int \D{t} \,
    \D{^2 x}  \, \left( \phi b - \epsilon_{ij} e_{jl} a_{li}\right)\\&+ \frac{1}{2\pi}   \int \D{t} \, \D{^2x}  \,\left( \Phi b - \epsilon_{ij} e_{jl} A_{li}\right)
\end{align}
which, upon integration of the dynamical gauge fields $(\phi,a_{ij})$, results in an action for the background gauge fields
\begin{equation}\label{eqn:CS_backgroundfields}
    S_{\textup{BG}}[A_{ij}, \Phi]=-\frac{1}{4\pi k }\int \D{t}  \D{^2 x} \left( \Phi B - \epsilon_{ij} E_{jl} A_{li}\right)
\end{equation}
where the background electric $E_{ij}$ and magnetic $B$ fields are defined using $(\Phi,A_{ij})$ in a fashion similar to \eqnref{eqn:EB}.
Varying $S_{\textup{BG}}$ w.~r.~t.~ $\Phi$ and $A_{ij}$, we recover
\begin{equation}\label{eqn:hallresponse}
        \rho^u = \frac{1}{2\pi k} B, \;\;\;\;
        J^u_{ij} =\frac{1}{2\pi k} \frac{\epsilon_{il} E_{lj}+\epsilon_{jl} E_{li}}{2}
\end{equation}
i.~e., a fractional dipolar fractonic Hall effect reported in \cite{PremPretkoNandkishore}, whose Hall response is determined by the level $k$.

\begin{figure}
    \centering
    \includegraphics[width=\linewidth]{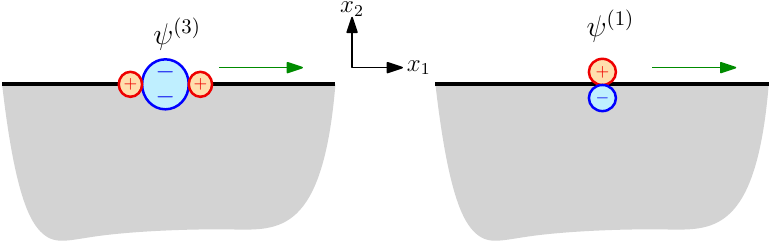}
    \centerline{(a)~~~~~~~~~~~~~~~~~~~~~~~~~~~~~~~~~~~~~~~~~~~~~~~~~~~~~~~~~~~~(b)}
    \caption{{\bf Edge Excitations:} Semi-infinite bulk $x_2 < 0$, terminated by edge $x_2 =0$, with $x_1$-axis tangent  and $x_2$-axis normal to the edge. (a) Shows the chiral fractonic edge modes where charges and longitudinal dipoles (moment along the tangent) are immobile, and the quadrupole shown (and higher multipoles) are mobile. (b) Shows the chiral non-fractonic edge mode with transverse dipoles (moment along the normal).   
    }
    \label{fig:edge}
\end{figure}

\paraheading{FCS Theory with an Edge} A crucial question to be addressed is the physics of fractonic theories in the presence of edges, i.~e., the spatial region on which the theory is defined possesses a boundary edge. The boundary edge is characterized by a one-dimensional curve with a coordinate $s$, a unit tangent vector $T_i(s)$, and a unit outward normal $N_i(s)$. Under a gauge transformation described by a function $f(t,x_1,x_2)$
\begin{equation}\label{eqn:GaugeTransBG}
    \begin{split}
       \Phi \to \Phi + \dou_t f, \;\;\; A_{ij} \to A_{ij} +\left( \dou_i \dou_j -\frac{1}{2} \dou^2 \delta_{ij} \right) f, \;\;\; 
        \end{split}
\end{equation}
the bulk does not contribute owing to the Bianchi identity, but the edge produces an anomaly (\SMRef{\secref{SM:sec:FCSBoundary}})
\begin{equation}
\begin{split}
     \delta S_{\textup{BG}}=-\frac{1}{4\pi k}\int  \D{t} \, \D{s} & \left( E_{jk} (T_j T_k  \dou_s f  + T_j N_k \dou_n f) \right. \\ 
     & \left. +  N_k N_j  \dou_s E_{jk}f - N_k T_j \dou_n E_{jk}f\right)
\end{split}
\end{equation}
where $\dou_s, \dou_n$ are tangential and normal derivatives along the boundary, and $E_{ij}$ is the background electric field. Specializing to a semi-infinite bulk with a straight boundary (see \figref{fig:edge}) where $s=x_1, T_i =  \delta_{i1}, N_i = \delta_{i2}$, we get 
\begin{equation}\label{eqn:BGAnomaly}
 \delta S_{\textup{BG}} =-\frac{1}{4\pi k}\int  \D{t} \D{x_1}  \left[\left(-2 \dou_1 E_{11} -\dou_2 E_{12}\right) f +  E_{12}\dou_2 f \right],
\end{equation}
as the gauge anomaly. This anomaly has to be compensated by some additional set(s) of degrees of freedom on the edge that also couple to the background gauge fields (with an effective action $S_{\textup{edge}}$) in an anomalous fashion, but in an opposite sense so that the overall system is gauge invariant\cite{Wen_IJMPB1992}. This allows us to write down the anomaly of additional edge degrees of freedom as
\beq
\delta S_{\textup{edge}} = \frac{1}{4\pi k}\int   \D{t} \D{x_1} \left[\left(-2 \dou_1 E_{11} -\dou_2 E_{12}\right) f +  E_{12}\dou_2 f \right]
\eeq
so that the total action $S_{\textup{BG}} + S_{\textup{edge}}$ has no gauge anomaly.   We explore the nature of these edge degrees of freedom, dubbed ``edge matter'', by observing that (\SMRef{\secref{SM:sec:FCSBoundary}}) the term $2 \dou_1 E_{11} +\dou_2 E_{12}$ in \eqnref{eqn:BGAnomaly} is the force on a quadrupole of with $\Theta = 2$ as shown in \figref{fig:edge}(a). Further, the term $E_{12}$ is the force on a unit $(D=1)$ dipole {\em transverse} to the edge as shown in  \figref{fig:edge}(b). These observations suggest that there are {\em two} distinct edge matter fields, one that is fractonic with {\em immobile charges, immobile longitudinal dipoles,} and {\em mobile higher multipoles}, and another non-fractonic field with {\em mobile transverse dipoles}. This conclusion is arrived at by noting that the quantity ${\cal E}^{(3)}=2 \dou_1 E_{11} +\dou_2 E_{12} = \dou_1^3 \Phi - \dou_t \left( 2 \dou_1 A_{11} +\dou_2 A_{12} \right)$ establishing that the quantity ${\cal E}^{(3)}$ is a {\em third} rank electric field defined on the edge via the edge gauge fields $(\Phi^{(3)}, A^{(3)})$, where $\Phi^{(3)}(t, x_1) = \Phi(t, x_1,x_2=0)$ and $A^{(3)} = \left( 2 \dou_1 A_{11} +\dou_2 A_{12} \right)$ evaluated on the edge. Under a gauge transformation, $(\Phi^{(3)},  A^{(3)}) \to (\Phi^{(3)} + \dou_t f, A^{(3)} + \dou^3_1 f)$, confirming the rank-3 nature of these edge gauge fields. Thus, we see that the first type of edge matter coupling to $(\Phi^{(3)}, A^{(3)})$ is expected to be fractonic with { immobile charges, longitudinal dipoles} (dipole moment along the edge tangent), but mobile quadrupoles. Turning to the quantity ${\cal E}^{(1)}=E_{12} = \dou_1(\dou_2 \Phi) - \dou_t A_{12}$, is a first rank electric field defined from gauge fields $(\Phi^{(1)} = \dou_2 \Phi,  A^{(1)} = A_{12})$ (where the r.h.s. quantities are defined on the edge) which transform as $(\Phi^{(1)}, A^{(1)}) \to (\Phi^{(1)} + \dou_t (\dou_2 f),  A^{(1)} + \dou_1(\dou_2 f))$, and thus the second type edge matter is non-fractonic. In short, we see that gauge anomaly \eqnref{eqn:BGAnomaly} of the FCS theory arises from two separate contributions of the gauge transformation field $f$ (see \ref{eqn:GaugeTransBG}); first through $f$ defined on the boundary, and second through a normal derivative of $f$ defined on the boundary. For a general $f$, these two are ``independent parameters", i.~e., we can choose $f$ whose normal derivative vanishes, or we can choose an $f$ which vanishes on the boundary, but its normal derivative does not.

The discussion above allows us to write down the  effective action for the edge background gauge fields
\beq
\begin{split}
S_{\textup{edge}}= \sum_{n=1,3}\half & \int \D{^2 X} \D{^2 X'} {\cal A}^{(n)}_{\alpha}(X) {K^{(n)}}^{\alpha \beta}(X-X'){\cal A}^{(n)}_{\alpha} (X') 
\end{split}
\eeq
where $X = (t,x_1)$ is the space-time coordinate of the edge, ${\cal A}^{(n)}_{\alpha} \equiv (\Phi^{(n)}, A^{(n)}),\,n=1,3$ defined in the previous paragraph, and $\alpha,\beta$ run over $t$ and $x_1$, ${K^{(3)}}^{\alpha \beta}(X-X') =\ci \mean{{j^{(3)}}^\alpha(X){j^{(3)}}^\beta(X')}$ is the response determined by the current (${j^{(3)}}^\alpha$) correlation function of fractonic edge matter that couples to ${\cal A}^{(3)}_{\alpha}$. Similarly, ${K^{(1)}}^{\alpha,\beta}(X,X') =\ci\mean{{j^{(1)}}^\alpha(X){j^{(1)}}^\beta(X')}$, where ${j^{(1)}}^\alpha$ is the current of the non-fractonic edge matter that couples to ${\cal A}^{(1)}_{\alpha}$. 

The action $S_{\textup{BG}}+S_{\textup{edge}}$ can be made anomaly-free (\SMRef{\secref{SM:sec:FCSBoundary}}) if the currents ${j^{(n)}}^{\alpha}$ ($\alpha=0 \equiv t, \alpha=1\equiv x_1$) satisfy the following algebra
\beq\label{eqn:KacMoody}
\begin{split}
\mean{[j_q^{(n)+}, j_{q'}^{(n)+}]} & = -\frac{q^n}{2 \pi k} \delta_{q+q',0}, \;\; n=1,3 \\
\mean{[j_q^{(n)+}, j_{q'}^{(n)-}]} & = 0 \\
\mean{[j_q^{(n)-}, j_{q'}^{(n)-}]} & = 0 
\end{split}
\eeq
where $j^{(1)\pm} = \half ({j^{(1)}}^0 \pm \frac{1}{v_1} {j^{(1)}}^1), j^{(3)\pm} = \half ({j^{(3)}}^0 \mp \frac{1}{v_3} {j^{(3)}}^1)$, $q$ is a wavevector, $j^{(n)}_q$ is the Fourier transform of $j^{(n)}$, and $v_n$ is a characteristic ``velocity'' of the edge excitations $\omega \sim v_n q^n$ for the fields labeled $n=1$ and $n=3$. For the edge matter that couples to ${\cal A}^{(1)}_\alpha$, \eqnref{eqn:KacMoody} is the usual U(1) Kac-Moody algebra\cite{Wen_PRL1990, Wen_PRB1991}, while for the fractonic edge matter with $n=3$, we have obtained a new algebra \eqnref{eqn:KacMoody} with $n=3$, which we dub the U(1) {\em fractonic} Kac-Moody algebra.

\paraheading{Edge Excitations} The above construction enables us to explicitly uncover the nature of the edge excitations by finding representations of the Kac-Moody algebras. The $n=1$ non-fractonic edge excitations corresponding to the transverse dipoles (see \figref{fig:edge}) can be described by a chiral boson\cite{Wen_PRB1991} field $\psi^{(1)}(t,x)$ which disperses as $\omega \sim v_1 q$, where $v_1$ is a non-universal velocity introduced above. Turning now to the representations of the fractonic algebra \eqnref{eqn:KacMoody} with $n=3$,  we observe that since these currents couple to rank-3 gauge fields in $1+1$ dimensions and $j^{(3)}$ currents must arise from fractonic matter with charge, dipole moment and quadrupole moment conservation (classically). In line with this,  we introduce a bosonic field $\psi^{(3)}(t,x_1)$ and demand that the transformation $\psi^{(3)} \to \psi^{(3)} + a +b x_1 + c x_1^2$ is a symmetry of the action, where $a, b, c$ are arbitrary real numbers. We see that $\dou_1^3 \psi^{(3)}$ is the simplest term invariant under this transformation,  using which we write the action
\begin{equation}
    S_{\textup{edge}}^{(3)}= \frac{k}{4\pi}\int \D{t} \, \D{x_1}\,  \left(-\dou_t \psi^{(3)} \dou_1^3 \psi^{(3)} +v_3 \left(\dou_1^3 \psi^{(3)} \right)^2 \right)
\end{equation}
which is a fractonic chiral boson with a dispersion $\omega = v_3 q^3$.
The conjugate momentum to $\psi^{(3)}$ is $\Pi^{(3)}(x)=-\frac{k}{4\pi} \dou_1^3 \psi^{(3)}$. Quantizing this theory using the method of Dirac brackets results in $[\psi^{(3)}(x_1),\Pi^{(3)}(x_1')] = \frac{\ci}{2} \delta(x_1-x'_1)$, and
\begin{equation}\label{eqn:Psi3Commutation}
    [\psi^{(3)}(x_1), \psi^{(3)}(x'_1)]= \ci \frac{ \pi}{2k} \left(x_1-x'_1\right)^2 \textup{sgn} \left(x_1-x'_1\right)
\end{equation}
This relation can be used to obtain (\SMRef{\secref{SM:sec:FCSBoundary}})  ${j^{(3)}}^0 = -\frac{1}{v_3}{j^{(3)}}^1 = \frac{1}{2\pi} \dou_1^3 \psi^{(3)}$, which in turn satisfies the current algebra \eqnref{eqn:KacMoody} with $n=3$. Further (\SMRef{\secref{SM:sec:FCSBoundary}}), \eqnref{eqn:Psi3Commutation} allows us to conclude that  $e^{-\ci Q \ell^{-1} \psi^{(3)}\left(x_1\right)}$ creates a charge of strength $\frac{Q}{k} \ell^{-1}$, $e^{-\ci D \dou_{x_1} \psi\left(x_1\right)}$ creates a longitudinal dipole of strength $\frac{D}{k}$, and finally, $e^{\ci \Theta \ell \dou_{x_1}^2 \psi\left(x_1\right)}$ creates a quadrupole of strength $\frac{\Theta}{k} \ell$ at the point $x_1$ on the edge. All of these excitations are fractionally background gauge charged.

\begin{figure}
    \centering
    \includegraphics[width=\linewidth]{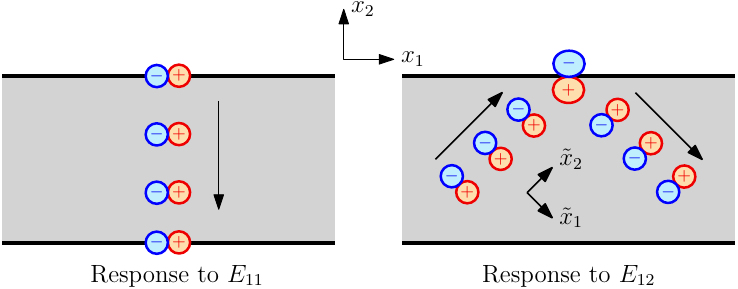}
    \centerline{(a)~~~~~~~~~~~~~~~~~~~~~~~~~~~~~~~~~~~~~~~~~~~~~~~~~~~~~~~~(b)}
    \caption{{\bf Bulk Response from Edge Anomaly:} (a) In the presence of a uniform $E_{11}$, the bulk current of longitudinal dipoles accounts for the anomaly in the longitudinal dipole density of the top and bottom edges.  (b) In the presence of a uniform $E_{12}$, anomaly of the transverse dipole density on edges is accounted for by the bulk flow of dipoles oriented at $\pm 45^{\circ}$ leading to the currents $\tilde{J}_{11} = \tilde{J}_{22} = 0, \tilde{J}_{12} = \tilde{J}_{21}$ in the frame $\tilde{x}_1-\tilde{x}_2$, consistent with $J_{11}$ predicted by the bulk response. }
    \label{fig:anomaly_transport}
\end{figure}

It is important to note the key differences between the usual (non-fracton) CS theory (such as that of the fractional Hall effect) and the FCS theory apropos the edge matter. In the usual CS theory with level $k$, the gauge anomaly can be canceled by a {\em single} non-fractonic chiral boson excitation. In contrast, the FCS theory requires {\em two} distinct edge excitations, one of which is fractonic and another non-fractonic.  The similarity of the two theories lies in the fact that the current algebras in both cases are determined by the level of the CS theory, which also determines the bulk response and the statistical phases of mobile bulk excitations. It is also interesting to elucidate how the two types of anomalous edge matter are related to the bulk response \eqnref{eqn:hallresponse}, which can be explicitly written as $J_{11} = \frac{1}{2 \pi k} E_{12}$ and $J_{12} = -\frac{1}{2 \pi k} E_{11}$. Considering a system with only $E_{11} \ne 0$, we see that the rate of change of the longitudinal dipole density on the edge inferred from the anomaly (accounting for covariant anomalies\cite{Stone1991,Stone2012})   of the fractonic edge mode on the top edge is $-\frac{1}{\pi k} E_{11}$, while that on the bottom edge is $\frac{1}{\pi k} E_{11}$ (\SMRef{\secref{SM:sec:fractonicAnomalyinflow}}).  This implies the presence of a flux $\Sigma = -\frac{1}{\pi k} E_{11}$ of longitudinal dipoles normal to the top edge; the flux is $-\Sigma$ on the bottom edge. This flux implies the presence of a bulk current of longitudinal dipoles (see \figref{fig:anomaly_transport}(a)) with $J_{11}=J_{22}= 0, J_{12}= J_{21} = \frac{\Sigma}{2}= -\frac{1}{2 \pi k} E_{11}$, which illustrates the relationship between the anomaly of the fractonic edge mode and the bulk Hall response. Further, in the case when only $E_{12} \ne 0$, the anomaly of the non-fractonic modes comprising of transverse dipoles, gives the rate of change of a transverse dipole density on the top(bottom) edges as $\mp\frac{1}{2 \pi k} E_{12}$. This change of transverse dipole densities on the edges can accomplished by the transport of dipoles oriented at $\pm 45^\circ$ as shown in \figref{fig:anomaly_transport}(b), leading precisely to the $J_{11}$ predicted by the bulk response(\SMRef{\secref{SM:sec:fractonicAnomalyinflow} for details).}

\begin{figure}
    \centering
    \includegraphics[width=\linewidth]{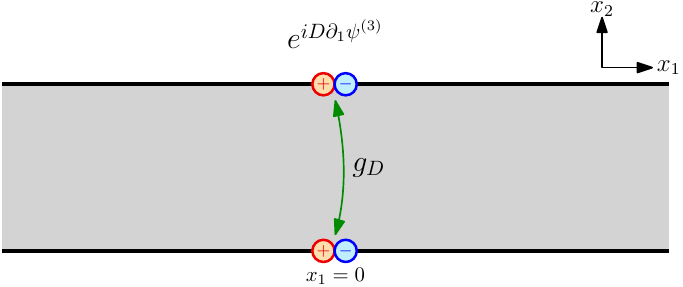}
    \caption{{\bf Allowed Tunneling Processes Between Edges:} Longitudinal dipoles of strength $D$ are allowed to tunnel from a point ($x_1=0$) the top $(+)$-edge to the bottom $(-)$-edge with amplitude $g_D$. Immobile charges and transverse dipoles cannot tunnel. }
    \label{fig:edge_stability}
\end{figure}

\paraheading{Edge-to-Edge Tunneling} Finally, we explore the physics of edge-to-edge tunneling. In the context of the fractional Hall effect, such physics was studied in \cite{MoonKaneGirvinFischer1993}, who introduced local tunneling between the two edges of a sample. Exploring an analogous scenario, we study a system with two parallel edges. Unlike in the fractional Hall system, not all edge excitations can tunnel from one edge to the other owing to the fractonic constraints of the theory. In particular, tunneling of charges, or transverse dipoles, is forbidden. The (edge-immobile) longitudinal dipole as shown in \figref{fig:edge_stability} is the lowest order multipole that can tunnel from one edge to the other (\SMRef{\secref{SM:sec:StatbilityRG}} for details). We study tunneling of such dipoles at a single point $(x_1=0)$ on the edge. 
The quantity $g_D$ determines the amplitude of the tunneling of a dipole of strength $D$ ($D/k$ relative to the background gauge fields) at the point $x_1=0$ from the top edge to the bottom edge, and all integer values of $D$ are allowed.  Exploring the flow of $g_D$ using the renormalization group, we find (\SMRef{\secref{SM:sec:StatbilityRG}})
\beq
    \frac{\D{g_D}}{\D{l}} = \left(3-\frac{D^2}{\abs{k}} \right) g_D.
\eeq
We thus show that the tunneling of all dipoles of strength $D^2 < 3 \abs{k}$ is relevant. In particular, $D=1$ dipole will lead to a growth of $g_1$ for {\em any} value of $k$. This may suggest that tunneling of longitudinal dipoles from one edge to the other will produce edge deformation owing to nonperturbative effects at the large values of $g_D$. This is to be compared with  the usual non-fractonic CS theory (e.~g., fractional quantum Hall (FQH) system \cite{MoonKaneGirvinFischer1993}).

\paraheading{Discussion} Adopting an effective description using a second-rank fractonic Chern-Simons (FCS) theory, we have revealed the rich and intriguing edge physics of fractonic systems, connecting it to the braiding statistics of bulk excitations elucidated here, and the previously known bulk transport properties.  The key feature is that even in the simplest realization of such a system with boundaries, there are {\em two} types of chiral edge excitations, one of which is fractonic and the other non-fractonic. The properties of the edge excitations (i.~e., the current algebra they satisfy) are determined by the level $k$ of the FCS theory, which also determines the transport properties and statistical phase of braiding of bulk quasiparticles.

Many of the features found in the second-rank FCS considered here also hold for higher-rank FCS theories; generically, in a rank $N$ trace-free bulk theory, we surmise the presence of $N$ gapless modes on the edge. Furthermore, we note in the type of straight edges considered in this paper (see \figref{fig:edge}) that the fractonic and non-fractonic edge modes do not interact with each other. This is no longer true when we consider a curved edge/boundary; this aspect is notably different from the non-fractonic Chern-Simons theory where the curvature of the edge plays no role. Other interesting directions for further study include the exploration of gappability of these fractonic edge theories in the presence of multiple higher-rank gauge fields similar to that in the usual Chern-Simons theory\cite{Ganeshan2022}. Generalizations to situations with vector fractonic charges\cite{Seiberg2020,Manoj2021} will also be interesting.

We expect the results uncovered here to hold in higher dimensions as well, since the existence of multiple edge modes in fractonic systems arises from the larger freedom in defining gauge transformations in higher-rank theories.

Even though experiments on fractonic systems may be challenging in the short term, we conclude the paper by noting that our results offer opportunities for future experiments. 
A recent experiment on an FQH platform \cite{Ruelle2024} introduces engineered edge excitations which are produced through time-resolved gate pulses. That approach might be further generalized to allow the generation of certain types of edge excitations in the present case of fractonic edges.

\noindent
{\em Acknowledgement:} BPP thanks PMRF program of the Ministry of Education, India. 
Y.G. acknowledges support by grant no 2022391 from the United States - Israel Binational Science Foundation
(BSF), Jerusalem, Israel, and the Minerva foundation. Y. G. is an incumbent of InfoSys chair at IISc Bangalore, and Deutsche Forschungsge-meinschaft (DFG) through grant No. MI 658/10-2.
VBS thanks SERB, DST, India for support. The authors thank Nandagopal Manoj for an insightful discussion.

\bibliography{ref}


\ifdefined\makeSM

\clearpage
\newpage
\appendix

\renewcommand{\appendixname}{}
\renewcommand{\thesection}{{S\arabic{section}}}
\renewcommand{\thefigure}{S\arabic{figure}}
\renewcommand{\theequation}{\thesection.\arabic{equation}}
 
\setcounter{page}{1}
\setcounter{figure}{0}

\begin{widetext}

\maketitle

\centerline{\bf Supplemental Material}
\medskip
\centerline{for}
\medskip
\centerline{\large{\bf \mytitle}}
\medskip
\centerline{by \authorZero, \authorOne, and \authorTwo}
\bigskip
\end{widetext}

\clearpage
\newpage

\appendix

\bigskip

	\paraheading{\bf Conventions used} For the discussions below in Sections \ref{SM:sec:chargeQuantization}, \ref{SM:sec:quantk}, we consider the fractonic Chern-Simons theory on the spacetime $ S^1 \times T^2$. $S^1$ is the compactified (imaginary) time  direction and $T^2$ is the spatial torus. We denote a point in the spacetime as $(\tau,x_1, x_2) = (\tau,\bx)$. $t$ ranges from $0$ to $\beta$ where $\beta$ is the inverse temperature. Torus $T^2$ is an (intrinsically) flat space, and we choose the metric tensor $g_{ij} = \delta_{ij}$ locally everywhere on $T^2$.
	
	 The convention adopted in Section \ref{SM:sec:FCSBoundary} for the metric tensor and the Fourier transform when studying the one-dimensional edge $x_2=0$ of length $L$ labeled by spacetime point $X=\left(t,x_1\right)$ is as follows. Metric is $\begin{pmatrix}
		-1 & 0 \\ 0&1 
	\end{pmatrix}$ so that $X^\mu=\left(t,x_1\right)$ and $X_\mu = \left(-t,x_1\right)$. The Fourier space is labeled by $P^\mu = \left(\omega, q\right)$, $q \in \frac{2 \pi}{L} \Integers$, and a function $g\left(X\right)$ is related to its Fourier transform $g\left(P\right)$ as
	\begin{equation}
		g(X) = \frac{1}{2\pi L} \int \D{\omega} \sum_q e^{\ci P_\mu X^\mu} g(P)
	\end{equation}
	
	\section{Quantization of charges and multipoles}\label{SM:sec:chargeQuantization}
	
	In this section, we discuss the quantization of charges in the effective U$(1)$ tensor gauge theory introduced in the main text. Further, appealing to the lattice geometry that underlies the effective theory, we discuss the quantization of dipoles and quadrupoles.
	
	Consider the U$(1)$ gauge fields $(\phi, a_{ij})$ where $a_{ij}$ is a traceless tensor. The dimensions of the gauge fields are $[a_{ij}]= [L]^{-1}$ and $[\phi]=[L][T]^{-1}$. Also, the dimension of the charge density $\rho$ coupling to these gauge fields is $[\rho]=[L]^{-3}$ (\eqnref{eqn:FCSaction}, main text). Equivalently, the dimension of a charge is $[L]^{-1}$, while a dipole moment is dimensionless. We write the charge as $Q \ell^{-1}$ where $Q$ is a dimensionless number and $\ell$ is a microscopic length scale such as the lattice spacing.
	
	Consider the Wilson loop operator $W^Q(x_1, x_2)$ associated with a charge $Q \ell^{-1}$:
	\begin{equation}\label{SM:eqn:Wilson}
		W^Q(x_1, x_2)=\exp{\left(i Q \ell^{-1} \int \limits_{0}^{\beta} \D{\tau} \, \phi(\tau, x_1, x_2) \right)}
	\end{equation} 
	For the above operator to be gauge invariant under the large gauge transformation $\phi \to \phi + \dou_\tau \xi$, where $\xi(\tau,x_1,x_2) = \frac{2\pi \ell}{\beta} \tau$, the number $Q$ must be an integer.
	
	 When the charges are placed on a lattice, the allowed values of the dipole moment are $\bD=\ell^{-1} \sum_I Q_I \bx_I$, with $\sum_I Q_I=0$. Here $\bx_I$ is the position vector of the lattice site labeled by $I$.  
  The choice of $\ell$ as the lattice spacing is natural as it renders the magnitude of the elementary dipole moment to be unity. Larger dipoles can be constructed from these elementary dipoles.
	 \begin{figure}
	     \centering
	     \includegraphics[width=0.4\linewidth]{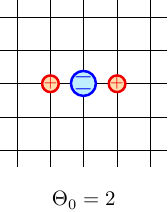}
	     \caption{$\Theta_0$ for ultraviolet realization on a square lattice ($\Theta_0=2$). 
      }
	     \label{SM:fig:UVlattice}
	 \end{figure}
	 
  Similarly, an allowed value of the trace the quadrupole moment tensor $\Theta_{ij}= \ell^{-1} \sum_I Q_I (x_I)_i (x_I)_j$ (with $\sum_I Q_I=0$ and $\sum_I Q_I \bx_I=0$) is of the form $\Theta \ell$, where $\Theta$ is an integer. For the square lattice, $\Theta=0,2,-2,4,\dots$ are allowed (see \figref{SM:fig:UVlattice}). 
  It will be useful to define $\Theta_0$ as the GCD (greatest common divisor) of all the allowed values of $\Theta$. For the square lattice, $\Theta_0=2$.

	 \section{Lagrangian of a quadrupole}\label{SM:sec:quadrupoleLagrange}
	 In the tensor gauge theory considered in the main text, a point quadrupole is the simplest multipole that has an unconstrained mobility. In this section, we derive the action for a quadrupole $\Theta_{ij}$ moving under the influence of tensor gauge fields. Using this, we derive the Aharonov-Bohm phase seen by a quadrupole and show that the total magnetic flux on $T^2$ is quantized. We denote the trace of the quadrupole moment tensor $\Theta_{ii}= \ell \Theta$
	 
	 The charge density is $\rho= \half \Theta_{ij} \dou_i \dou_j \delta^{(2)}\left(\bx - \bx_0\left(t\right) \right)$ for a quadrupole $\Theta_{ij}$ moving along a path $\bx_0\left(t\right)$. Stating that quadrupoles are fully mobile is equivalent to saying that there is a rank$-2$ traceless current density $J_{ij}$ ($J_{ii}=0$) satisfying the continuity equation $\dou_t \rho + \dou_i \dou_j J_{ij} =0$ for \emph{any arbitrary} function $\bx_0\left(t\right)$. Let $\bv = \frac{\D{\bx_0\left(t\right)}}{\D{t}}$ be the velocity of the quadrupole. Then the current density $J_{ij}$ is
	 \begin{equation}\label{SM:eqn:quadrupoleCurrent}
  \begin{split}
	 	J_{ij} = &  \half \Theta_{ij} v_k \dou_k \delta^{(2)} \left(\bx -\bx_0\left(t\right)\right) \\
   & -\frac{1}{4} \ell \Theta v_k  \left(\epsilon_{ik} \epsilon_{jl} + \epsilon_{il} \epsilon_{jk} \right) \dou_l \delta^{(2)} \left(\bx -\bx_0\left(t\right)\right)
   \end{split}
	 \end{equation}
	 The second term above is added to make $J_{ij}$ traceless, while still satisfying the continuity equation. Now, assuming the usual non-relativistic kinetic energy for the motion of a quadrupole of unit mass, we can write the action for the quadrupole as
	 \begin{equation}\label{SM:eqn:quadrupoleActionDefn}
	 	S_{\text{quadrupole}}= \int \D{t} \half \dot{\bx}_0^2 + \int \D{^2 x} \D{t} \left( \rho \phi - J_{ij} a_{ij} \right).
	 \end{equation}
	 Plugging in the charge density $\rho$ and the current density $J_{ij}$ from \eqnref{SM:eqn:quadrupoleCurrent}, we find
	 \begin{equation}\label{SM:eqn:quadrupoleAction}
	 	S_{\text{quadrupole}}= \half \int \D{t} \, \dot{\bx}_0^2 +\Theta_{ij}\left(\dou_i \dou_j \phi + v_k \dou_k a_{ij}\right) +\ell \Theta v_i \dou_j a_{ij}.
	 \end{equation}
	 The Euler-Lagrange equation of motion reads
	 \begin{equation}\label{SM:eqn:forceQuadrupole}
	 	\ddot{x}_{0k} = \half \left(\Theta_{ij} \dou_k e_{ij}  + \ell \Theta \dou_i e_{ik}\right) + \half \ell \Theta \epsilon_{ki} v_i b
	 \end{equation}
where $e_{ij}$ and $b$ are respectively the electric and magnetic fields as defined in \eqnref{eqn:EB}.	
The right hand side of the above \eqnref{SM:eqn:forceQuadrupole} should be interpreted as the generalized Lorentz force on a quadrupole.
	 
	 \subsection{Aharonov-Bohm phase: quadrupole}\label{SM:sec:quadrupolePhase}
	 We will show that, in this tensor gauge theory, taking a quadrupole adiabatically around a closed path ${\cal C}$ in the presence of static magnetic field results in a U$(1)$ phase. For this calculation, assume that the gauge field $a_{ij}$ is time-independent.
	 
	 The phase can be obtained by inspecting the term in the action that is invariant under reparameterization of the path. i.~e., a term that depends only on the path taken in real space, but not on the details of the speed along the path. In this case, this term can be read off to be $\half \oint_{{\cal C}} \D{x_{0k}} \, \left( \Theta_{ij} \dou_k a_{ij} + \ell \Theta \dou_j a_{kj}\right)$. The first term drops out because of the integration over a closed path, and using the Stokes' theorem, the second term simplifies to $\half \ell \Theta \int_{{\cal S}} \D{^2 x} \, \epsilon_{lk} \dou_l \dou_j a_{kj} = \half \ell \Theta \int_{{\cal S}} \D{^2x} \, b$. Here, ${\cal S}$ is a region enclosing the path ${\cal C}$. Hence, the Aharomov-Bohm phase in this case is
	 \begin{equation}\label{SM:eqn:quadrupolePhase}
	 	\exp{\left(\ci \frac{\ell \Theta}{2} \int_{{\cal S}} \D{^2x} \, b \right)}.
	 \end{equation} 
	 
	 \subsection{Quantization of magnetic flux}\label{SM:sec:fluxQuantization}
	 Following the familiar Dirac argument, we will show that the total magnetic flux on $T^2$, a closed surface, is quantized in this tensor gauge theory.
	 
	 The Aharonov-Bohm phase associated with a quadrupole \eqnref{SM:eqn:quadrupolePhase} calculated using the region ${\cal S}$ should be same as the one calculated using its complement region $\overline{{\cal S}}$ in $T^2$. Hence, we must have $\exp{\left(\ci \frac{\ell \Theta}{2} \int_{T^2} \D{^2x} \, b  \right)}=1$ for any value of $\Theta$ allowed on the lattice. Thus, the magnetic flux has a minimum allowed value $\frac{4\pi}{\ell \Theta_0}$ where $\Theta_0$ is the GCD of the allowed values of $\Theta$ (\secref{SM:sec:chargeQuantization}). The quantization of magnetic flux then reads
	 \begin{equation}
	 	\ell \frac{\Theta_0}{4\pi}\int_{T^2} \D{^2x} \, b  \in  \mathbb{Z}.
	 \end{equation}

\section{Quantization of the fractonic Chern-Simons theory}\label{SM:sec:quantk}
First, we show that the level $k$ of the  fractonic Chern-Simons action \eqnref{eqn:FCSaction} is quantized to be an integer, assuming that the underlying UV has the square lattice geometry.

The FCS action in  \eqnref{eqn:FCSaction} can also be written as 
\begin{equation}
	S=\frac{k}{4\pi} \int \D{\tau} \D{^2 x} \left(2 \phi b + \epsilon_{jk} a_{ij} \dou_t a_{ki} \right).
\end{equation}
Now consider a gauge field configuration ($\phi$, $a_{ij}$) so that the total magnetic flux is $ \int_{T^2} \D{^2x} \, b = \frac{4\pi}{\ell \Theta_0}$. The change in the action under a large transformation $\phi \to \phi + \frac{2\pi \ell}{\beta}$ is
\begin{equation}
	\begin{split}
		\delta S&= \frac{k}{4\pi}\int \D{\tau} \D{^2 x}  \left(2\times \frac{2\pi \ell }{\beta} b \right)\\
		&=\frac{4\pi k}{\Theta_0}  \in 2\pi \mathbb{Z}.
	\end{split}
\end{equation}
We demand that the $\delta S$ is an integer multiple of $2\pi$ for gauge invariance. For the square lattice, $\Theta_{0}=2$, hence we have
\begin{equation}
	k \in \mathbb{Z}.
\end{equation}
Now, we discuss the quantization of the fractonic Chern-Simons theory \eqnref{eqn:FCSaction}. Note that $a_{11}$ and $a_{12}$ are the independent parameters of the traceless tensor $a_{ij}$. The momenta conjugate to these fields are $\Pi_{11}(\bx)=-\frac{k}{2\pi} a_{12}(\bx)$, $\Pi_{12}(\bx)=\frac{k}{2\pi} a_{11}(\bx)$. We find that the commutators are
\begin{equation}\label{SM:eqn:CSCommutators}
	\begin{split}
		[a_{11}(\bx),a_{11}(\bx')]&= [a_{12}(\bx),a_{12}(\bx')]=0 \\
		[a_{11}(\bx),a_{12}(\bx')] &= -\ci \frac{\pi}{k} \delta^{(2)}(\bx-\bx')
	\end{split}
\end{equation}

\section{Statistical Phase of Dipoles}\label{SM:sec:dipoleStatPhase}
In this section, we show that the dipole excitations of the fractonic Chern-Simons theory have nontrivial braiding statisitcs. A dipole in this theory is mobile only along the direction transverse to its dipole moment (lineon). To compute the braiding phase of dipoles $\bD$ and $\bD'$, we consider the process shown in \figref{fig:stat_phase}, main text.

The operator that transports the dipole $\bD$ from $1$ to $2$ is $W= \exp{\left(\ci \int\limits_{1}^2 \D{x_i} \, a_{ij}(\bx) D_j \right)}$. Similarly, the operator transporting the dipole $\bD'$ from $1'$ to $2'$ is $W' =\exp{\left(\ci \int\limits_{1'}^{2'} \D{x}_i' \, a_{ij}(\bx') D'_j \right)} $. The process of braiding is captured by the operator $\left(W'\right)^{-1} W^{-1} W' W$, which is a U$(1)$ phase, as shown below.

Let $\alpha$ be the angle between the vectors $\bD$ and $\bD'$, and without loss of generality, let $\bD =D \left(1,0\right)$, $\bD' =D'\left(\cos \alpha, \sin \alpha\right)$. Assume that the the lines $1 \to 2$ and $1' \to 2'$ intersect at the origin $(0,0)$. A parameterization of the path $ 1 \to 2$ is  $\bx(s) = s(0,1)$, $-1<s<1$, and the path $1' \to 2'$ is parameterized by $\bx'(s) = s\left(\sin{\alpha},-\cos{\alpha}\right)$, $-1<s<1$. Using the Baker-Campbell-Hausdorff formula and the commutation relations \eqnref{SM:eqn:CSCommutators}, the operator performing the braiding $\left(W'\right)^{-1} W^{-1} W' W=$
\begin{equation}
	\begin{split}
	&\exp\left(\int \limits_{1}^2 \D{x}_i \int \limits_{1'}^{2'} \D{x}_k'  \,\,[a_{ij}\left(\bx\right), a_{kl}\left(\bx'\right)] D_j D_l \right) \\
	&= \exp\left(\ci \frac{\pi}{k} D D' \sin\left(2\alpha\right) \int \limits_{-1}^1 \D{s}\int \limits_{-1}^1 \D{s}' \delta\left(s \sin \alpha\right)  \delta\left(s'+s\cos \alpha \right)\right) \\
	&= \exp \left(\ci \frac{2\pi}{k} \bD \cdot \bD'\right).
	\end{split}
\end{equation}

\section{Fractonic Chern-Simons theory in a system with a boundary}\label{SM:sec:FCSBoundary}
 We consider the fracton Chern-Simons theory on a system with a boundary. The action for the background gauge fields \eqnref{eqn:CS_backgroundfields} is not gauge invariant when the region of space is truncated with a boundary. This leads to the emergence of gapless degrees of freedom on the edge. In this section, we give the details of the gauge anomaly, characterize the anomaly of the edge modes, derive the current algebra satisfied by these edge degrees of freedom, and find the theories that realize this algebra.
 
 Let the edge of the system be parameterized by a distance parameter $s$, and denote the unit tangent and outward normal vectors to the edge as $T_i(s)$  and $N_i(s)$, respectively. The change in the action \eqnref{eqn:CS_backgroundfields} under a gauge transformation
 \begin{equation}
 	\begin{split}
 		A_{ij} &\to A_{ij} +\left( \dou_i \dou_j -\frac{1}{2} \dou^2 \delta_{ij} \right) f \\
 		\Phi &\to \Phi + \dou_t f.
 	\end{split}
 \end{equation}
Since the bulk term vanishes due to the Bianchi identity, the surface term is
\begin{equation}
	\begin{split}
		\delta S_{\textup{BG}}&=  -\frac{1}{4\pi k} \int \D{t} \D{^2 x} \left(\dou_t(f B) - \dou_i(\epsilon_{ij} E_{jk} \dou_k f) +\dou_k (f \epsilon_{ij} \dou_i E_{jk} ) \right) \\
		&=-\frac{1}{4\pi k }\int  \D{t} \D{s} \left( -N_i \epsilon_{ij} E_{jk} \dou_k f + N_k f \epsilon_{ij} \dou_i E_{jk} \right)
	\end{split}
\end{equation}
	Denoting $\dou_s$ and  $\dou_n$ as the tangential and normal derivatives on the boundary, and assuming that the boundary is a straight line (as in \figref{fig:edge}),
	\begin{equation}
 \begin{split}
		\delta S_{\textup{BG}} = 
   -\frac{1}{4\pi k} \int  \D{t} \D{s}  & \left[\left(-2 T_i  T_j \dou_s E_{ij} -T_i  N_j \dou_n E_{ij}\right) f  \right. \\ 
   & \left. + \left( T_i  N_j E_{ij}\right) \dou_n f \right].
  \end{split}
	\end{equation}
Taking the boundary (edge) to be the $x_1$-axis i.~e., $x_2=0$,
\begin{equation}\label{SM:eqn:deltaSBGXaxis}
	\delta S_{\textup{BG}} =-\frac{1}{4\pi k}\int  \D{t} \D{x_1} \left[\left(-2 \dou_1 E_{11} -\dou_2 E_{12}\right) f +  E_{12}\dou_2 f \right].
\end{equation}
This gauge anomaly must be canceled by the anomaly of the edge modes to restore gauge invariance \cite{Wen_IJMPB1992}. The anomaly of these edge modes is then
\begin{equation}\label{SM:eqn:deltaSedge}
	\delta S_{\text{edge}} = \frac{1}{4\pi k}\int  \D{t} \D{x_1} \left[\left(-2 \dou_1 E_{11} -\dou_2 E_{12}\right) f +  E_{12}\dou_2 f \right].
\end{equation}
Now, let us look at the quantities in the above \eqnref{SM:eqn:deltaSBGXaxis} more carefully: The quantity ${\cal{E}}^{(3)}\left(t,x_1\right) \coloneqq 2 \dou_1 E_{11}\left(t,x_1,x_2=0\right) +\dou_2 E_{12} \left(t,x_1,x_2=0\right)$ defined on the edge is the component of force  (\eqnref{SM:eqn:forceQuadrupole}) along the edge $x_2=0$ on a quadrupole $\Theta_{ij}=\begin{pmatrix}
	2 & 0 \\
	0 & 0
\end{pmatrix}$ restricted to move on the edge. It can be written as ${\cal{E}}^{(3)} = \dou_1^3 \Phi^{(3)} - \dou_t A^{(3)}$, where $\Phi^{(3)} \left(t,x_1\right) \coloneqq \Phi\left(t,x_1,x_2=0\right)$, $A^{(3)}\left(t,x_1\right) \coloneqq 2\dou_1 A_{11}\left(t,x_1,x_2=0\right) +\dou_2 A_{12}\left(t,x_1,x_2=0\right)$. Under a gauge transformation, $A^{(3)}$ transforms as $A^{(3)} \to A^{(3)} + \dou_1^3 f$. Thus, we see that ${\cal{E}}^{(3)}$ is a rank-3 electric field emerging on the edge with gauge fields $\left(\Phi^{(3)}, A^{(3)}\right)$.

Similarly, ${\cal{E}}^{(1)}\left(t,x_1\right) \coloneqq E_{12} \left(t,x_1,x_2=0\right) $ is the component of force on a transverse dipole $\bD=\left(0,1\right)$ moving on the edge $x_2=0$. The relation ${\cal{E}}^{(1)} = \dou_1\Phi^{(1)} - \dou_t A^{(1)}$ suggests that ${\cal{E}}^{(1)}$ is a rank$-1$ gauge field on the edge with the gauge fields $\left(\Phi^{(1)}, A^{(1)}\right)$ where $\Phi^{(1)} \left(t,x_1\right) =\dou_2\Phi\left(t,x_1,x_2=0\right)$, $A^{(1)}\left(t,x_1\right) = A_{12}\left(t,x_1,x_2=0\right)$. We summarize the details of the emergent rank$-1$ and rank$-3$ gauge fields on the edge in \tabref{SM:tab:edgeGaugefields}.
\begin{table}
\begin{tabular}{|c|c|c|c|}
	\hline 	
	Rank & Gauge fields & Gauge transformation & Electric field\\
	\hline
	1 & \begin{tabular}{c}$\left(\Phi^{(1)}, A^{(1)}\right)$ \\ $\Phi^{(1)} = \dou_2 \Phi$ \\ $A^{(1)} = A_{12}$ \end{tabular} &  \begin{tabular}{c} $\Phi^{(1)} \to \Phi^{(1)}+ \dou_t \left(\dou_2 f\right)$ \\ $A^{(1)} \to A^{(1)}+ \dou_1 \left(\dou_2 f\right)$ \end{tabular} & ${\cal{E}}^{(1)} = \dou_1\Phi^{(1)} - \dou_t A^{(1)}$ \\
	\hline
	3 & \begin{tabular}{c} $\left(\Phi^{(3)}, A^{(3)}\right)$ \\ $\Phi^{(3)} = \Phi$ \\ $A^{(3)} = 2 \dou_1 A_{11} + $\\ $\dou_2 A_{12}$ \end{tabular} & \begin{tabular}{c} $\Phi^{(3)} \to \Phi^{(3)}+ \dou_t f$ \\ $A^{(3)} \to A^{(3)}+ \dou_1^3 f$ \end{tabular} & ${\cal{E}}^{(3)} = \dou_1^3\Phi^{(3)} - \dou_t A^{(3)}$ \\
	\hline
\end{tabular}
\caption{Table showing the gauge fields emergent on the edge and their gauge transformations.}
\label{SM:tab:edgeGaugefields}
\end{table}
Thus, we must have two distinct edge modes which couple to rank$-1$ and rank$-3$ gauge fields to cancel the gauge anomaly. Based on the discussion of last paragraph, the edge modes that couple to the rank$-1$  gauge fields are non-fractonic, and correspond to transverse dipoles which can move freely on the one-dimensional edge. Whereas, the edge modes that couple to rank$-3$ gauge fields are fractonic, with quadrupoles being the simplest mobile point excitations.

We now introduce some notation to write the equations more compactly. We denote the gauge fields on the edge as ${\cal A}^{(n)}_{\alpha} = \left(\Phi^{(n)}, A^{(n)}\right)$, where $n=1,3$ denotes the rank of the gauge fields. The gauge transformations acting on these gauge fields can be written compactly by introducing the symbol $w \left(n,\alpha\right)$ with $w\left(1,0\right)=w\left(1,1\right) =w\left(3,0\right)=1, w\left(3,1\right)=3$:
\begin{equation}\label{SM:eqn:edgeGaugetrans}
	{\cal A}^{(n)}_\alpha \to {\cal A}^{(n)}_\alpha  + \dou_\alpha^{w\left(n,\alpha\right)} {\cal F}^{(n)}.
\end{equation}
Here, ${\cal F}^{(1)}=\dou_2 f, {\cal F}^{(3)} = f$. In this notation, the rank$-n$ electric field on the edge can be written as ${\cal E}^{(n)} = - \epsilon^{\alpha \beta} \dou_\alpha^{w(n,\alpha)} {\cal A}^{(n)}_\beta$ and the edge anomaly \eqnref{SM:eqn:deltaSedge} can be written as
\begin{equation}\label{SM:eqn:deltaSedgecompact}
		\delta S_{\text{edge}} = \frac{1}{4\pi k}\int  \D{t} \D{x_1} \left(-{\cal E}^{(3)} {\cal F}^{(3)} +  {\cal E}^{(1)} {\cal F}^{(1)} \right).
\end{equation}
Written in terms the Fourier space variable $P^\alpha=\left(\omega, q\right)$, this becomes
\begin{widetext}
\begin{equation}\label{SM:eqn:deltaSedgeMomentum}
	\begin{split}
	\delta S_{\text{edge}}& = \frac{1}{4\pi k} \frac{1}{2\pi L }   \int \D{\omega} \sum_q  \left(-{\cal E}^{(3)} \left(P\right) {\cal F}^{(3)} \left(-P\right) +  {\cal E}^{(1)}\left(P\right) {\cal F}^{(1)} \left(-P\right) \right) \\
	&= \frac{1}{4\pi k} \frac{1}{2\pi L }   \int \D{\omega} \sum_q  \Bigg[\epsilon^{\alpha \beta} \left(\ci P_\alpha \right)^{w\left(3,\alpha\right)} {\cal F}^{(3)} \left(-P\right) {\cal A}^{(3)}_\beta \left(P\right) \Bigg. \\
	& \;\;\;\;\;\;\;\;\;\;\;\;\;\;\;\;\;\;\;\;\;\;\;\;\;\;\;\;\Bigg. -\epsilon^{\alpha \beta} \left(\ci P_\alpha \right)^{w\left(1,\alpha\right)} {\cal F}^{(1)} \left(-P\right) {\cal A}^{(1)}_\beta \left(P\right) \Bigg].
\end{split}
\end{equation}
\end{widetext}
\subsection{Anomaly of the edge modes}
Now, we write an effective action for the external gauge fields coupling to the edge modes,
\begin{equation}\label{SM:eqn:edgeEffaction}
\begin{split}
	S_{\textup{edge}}= \half & \int \D{^2 X} \D{^2 X'} {\cal A}^{(3)}_{\alpha}(X) {K^{(3)}}^{\alpha \beta}(X-X'){\cal A}^{(3)}_{\beta} (X') \\
	+ & \half  \int \D{^2 X} \D{^2 X'} {\cal A}^{(1)}_{\alpha}(X) {K^{(1)}}^{\alpha \beta}(X-X'){\cal A}^{(1)}_{\beta} (X') 
\end{split}
\end{equation}
where $X=\left(t,x_1\right)$ is the spacetime point on the edge, ${K^{(3)}}^{\alpha \beta}$ is the time-ordered correlation function of the fractonic currents on the edge: ${K^{(3)}}^{\alpha \beta} \left(X -X'\right) = \ci \expect{{j^{(3)}}^\alpha\left(X\right) {j^{(3)}}^\beta \left(X'\right)}$, and ${K^{(1)}}^{\alpha \beta}=\ci \expect{{j^{(1)}}^\alpha\left(X\right) {j^{(1)}}^\beta \left(X'\right)}$ is the time-ordered correlation function of the non-fractonic currents (that of transverse dipoles) on the edge.
In Fourier space,
\begin{equation}\label{SM:eqn:edgeEffactionMomentum}
	S_{\text{edge}} = \frac{1}{ 2\pi  L} \sum_{n=1,3} \int \D{\omega} \sum_q {\cal A}^{(n)}_\alpha \left(P\right) {K^{(n)}}^{\alpha \beta} \left(-P\right) {\cal A}^{(n)}_\beta \left(-P\right).
\end{equation}
The Fourier transform of the correlation function ${K^{(n)}}^{\alpha \beta}\left(P\right)$ satisfies ${K^{(n)}}^{\alpha \beta}\left(P\right)={K^{(n)}}^{ \beta \alpha}\left(-P\right)={{K^{(n)}}^{\alpha \beta}}^\ast\left(-P\right)$. The change in effective action \eqnref{SM:eqn:edgeEffactionMomentum}   under the gauge transformation \eqnref{SM:eqn:edgeGaugetrans} is

 \begin{widetext}
\begin{equation}\label{SM:eqn:deltaSeffedge}
	\begin{split}
	\delta S_{\text{edge}} = & \frac{1}{2\pi  L} \sum_n \int \D{\omega} \sum_q \Bigg[\left(-\ci P_{\alpha} \right)^{w\left(n,\alpha\right)} {\cal F}^{(n)} \left(-P\right)  {K^{(n)}}^{\alpha \beta} \left(P\right)  {\cal A}^{(n)}_\beta \left(P\right) \Bigg. \\
	\Bigg. &+  \left(\ci P_\alpha\right)^{w\left(n,\alpha \right)} \left(-\ci P_\beta\right)^{w\left(n,\beta\right)}  {K^{(n)}}^{\alpha \beta}\left(-P\right){\cal F}^{(n)} \left(P\right)   {\cal F}^{(n)} \left(-P\right) \Bigg].
\end{split}
\end{equation}
\end{widetext}

We want the above change in action be equal to the desired in \eqnref{SM:eqn:deltaSedgeMomentum}. Comparing \eqnref{SM:eqn:deltaSedgeMomentum} and \eqnref{SM:eqn:deltaSeffedge} allows us to conclude that
\begin{equation}\label{SM:eqn:bothAnomalies}
	\begin{split}
		\left(-\ci P_\alpha\right)^{w\left(1,\alpha\right)} {K^{(1)}}^{\alpha \beta}\left(P\right) &= \frac{-1}{4\pi k} \epsilon^{\alpha \beta} \left(\ci P_\alpha \right)^{w\left(1,\alpha\right)} \\
			\left(-\ci P_\alpha\right)^{w\left(3,\alpha\right)} {K^{(3)}}^{\alpha \beta}\left(P\right) &= \frac{1}{4\pi k} \epsilon^{\alpha \beta} \left(\ci P_\alpha \right)^{w\left(3,\alpha\right)}.
	\end{split}
\end{equation}
These are the conditions to be satisfied by the current correlators of the edge modes, and they characterize the anomalies of the edge modes. These conditions ensure that the last term in \eqnref{SM:eqn:deltaSeffedge} vanishes. For the edge modes coupling to rank$-1$ gauge fields, $w\left(1,0\right)=w\left(1,1\right)=1$ allows us to write the above condition as
\begin{equation}\label{SM:eqn:rank1anomaly}
	P_\alpha {K^{(1)}}^{\alpha \beta}\left(P\right) = \frac{1}{4\pi k} \epsilon^{\alpha \beta} P_\alpha.
\end{equation}
Expanding the above equation,
\begin{equation}
	\begin{split}
		-\omega {K^{(1)}}^{00}\left(\omega, q\right) +q {K^{(1)}}^{10}\left(\omega, q\right) &=-\frac{q}{4\pi k} \\
		-\omega {K^{(1)}}^{01}\left(\omega, q\right) + q {K^{(1)}}^{11}\left(\omega, q\right) &=-\frac{\omega}{4\pi k}.
	\end{split}
\end{equation}
For the fractonic edge modes coupling to rank$-3$ gauge fields, the anomaly can be written as
\begin{equation}\label{SM:eqn:rank3anomaly}
	\begin{split}
		\omega {K^{(3)}}^{00}\left(\omega, q\right) +q^3 {K^{(3)}}^{10}\left(\omega, q\right) &=\frac{q^3}{4\pi k} \\
		\omega {K^{(3)}}^{01}\left(\omega, q\right) + q^3 {K^{(3)}}^{11}\left(\omega, q\right) &=-\frac{\omega}{4\pi k}.
 	\end{split}
\end{equation}
The anomaly can also be cast in the form of a violation of the continuity equations in the presence of background electric fields. This can be seen as follows. Firstly, the expectation value of the current is related to the current correlator as
\begin{equation}
	\expect{{j^{(n)}}^\alpha\left(X\right)} = \int \D{^2 X'} {K^{(n)}}^{\alpha \beta} \left(X-X'\right) {\cal A}^{(n)}_\beta \left(X'\right).
\end{equation}
Dropping the expectation symbol, this can be written in the Fourier space as \begin{equation}
	{j^{(n)}}^\alpha \left(P\right)= \frac{1}{\sqrt{2\pi L}}{K^{(n)}}^{\alpha \beta} \left(P\right) {\cal A}^{(n)}_\beta \left(P\right).
\end{equation}
Using \eqnref{SM:eqn:bothAnomalies}, we obtain the result
\begin{equation}
	\begin{split}
		-\ci P_\alpha  {j^{(1)}}^\alpha \left(P\right) &= \frac{1}{\sqrt{2\pi L}} \frac{1}{4\pi k} {\cal E}^{(1)} \left(P\right) \\
		\left(-\ci P_\alpha \right)^{w\left(3,\alpha\right)} {j^{(3)}}^\alpha \left(P\right) &= - \frac{1}{\sqrt{2\pi L}} \frac{1}{4\pi k} {\cal E}^{(3)} \left(P\right).
	\end{split}
\end{equation}
Transforming back to the time and position space,
\begin{equation}\label{SM:eqn:consistentAnomalyContinuity}
	\begin{split}
		\dou_t {j^{(1)}}^0 + \dou_1 {j^{(1)}}^1 &= -\frac{1}{4\pi k} {\cal E}^{(1)}\\
		\dou_t {j^{(3)}}^0 + \dou_1^3 {j^{(3)}}^1 &= \frac{1}{4\pi k} {\cal E}^{(3)}.
	\end{split}
\end{equation}
The current ${j^{(n)}}^\alpha$ is obtained from the variation of the edge effective action \eqnref{SM:eqn:edgeEffaction}, and the anomaly equation for this current \eqnref{SM:eqn:consistentAnomalyContinuity} is called the consistent anomaly \cite{Stone2012}. The current ${j^{(n)}}^\alpha$ is not gauge invariant, and transforms under the gauge transformation \eqnref{SM:eqn:edgeGaugetrans} as
\begin{equation}
    {j^{(n)}}^\alpha \left(P\right) \to {j^{(n)}}^\alpha \left(P\right) + \frac{1}{\sqrt{2\pi L}} {K^{(n)}}^{\alpha \beta} \left(P\right) \left(\ci P_\beta\right)^{w\left(n,\beta\right)} {\cal F}^{(n)}\left(P\right).
\end{equation}
Using \eqnref{SM:eqn:bothAnomalies}, this transformation can be written as
\begin{equation}
\begin{split}
    {j^{(1)}}^\alpha \left(P\right) &\to {j^{(1)}}^\alpha \left(P\right) - \frac{1}{\sqrt{2\pi L}}  \frac{1}{4\pi k} \ci \epsilon^{\alpha \beta}P_\beta {\cal F}^{(1)}\left(P\right) \\
    {j^{(3)}}^\alpha \left(P\right) &\to {j^{(3)}}^\alpha \left(P\right) + \frac{1}{\sqrt{2\pi L}}  \frac{1}{4\pi k} \epsilon^{\alpha \beta}\left(\ci P_\beta\right)^{w\left(3,\beta\right)} {\cal F}^{(3)}\left(P\right).
\end{split}
\end{equation}
The gauge invariant edge current can be obtained from the variation of the full effective action $S=S_{\text{BG}} + S_{\text{edge}}$, and including the boundary term arising from the Chern-Simons tern $S_{\text{BG}}$. Under a variation of the gauge fields $\Phi \to \Phi + \delta \Phi, A_{ij} \to A_{ij} + \delta A_{ij}$, the variation in the Chern-Simons action is
\begin{equation}
\begin{split}
    \delta S_{\text{BG}} = \int \D{t} \D{^2 x} &\, \left(\rho^u \delta \Phi - J_{ij}^u \delta A_{ij} \right) \\ +& \frac{1}{4\pi k} \int \D{t} \D{x_1} \left(\epsilon^{\alpha \beta} {\cal A}^{(1)}_\beta \delta {\cal A}^{(1)}_\alpha  - \epsilon^{\alpha \beta} {\cal A}^{(3)}_\beta \delta {\cal A}^{(3)}_\alpha \right)
    \end{split}
\end{equation}
The first term gives the bulk Hall response with $\rho^u, J^u_{ij}$ as in \eqnref{eqn:hallresponse}. The second term gives a contribution to the gauge invariant edge current ${{\overline{j}}^{(n)}}^\alpha$ defined as follows:
\begin{equation}
    \begin{split}
        {{\overline{j}}^{(1)}}^\alpha &= {j^{(1)}}^\alpha + \frac{1}{4\pi k} \epsilon^{\alpha \beta} {\cal A}^{(1)}_\beta \\
        {{\overline{j}}^{(3)}}^\alpha &= {j^{(3)}}^\alpha - \frac{1}{4\pi k} \epsilon^{\alpha \beta} {\cal A}^{(3)}_\beta
    \end{split}
\end{equation}
${{\overline{j}}^{(n)}}^\alpha$ is invariant under the gauge transformations \eqnref{SM:eqn:edgeGaugetrans} and its anomaly, called the covaraint is given by
\begin{equation}\label{SM:eqn:covariantAnomalyContinuity}
	\begin{split}
		\dou_t {{\overline{j}}^{(1)}}^0+ \dou_1 {{\overline{j}}^{(1)}}^1 &= -\frac{1}{2\pi k} {\cal E}^{(1)}\\
		\dou_t {{\overline{j}}^{(3)}}^0+ \dou_1^3 {{\overline{j}}^{(3)}}^1&= \frac{1}{2\pi k} {\cal E}^{(3)}.
	\end{split}
\end{equation}
We discuss the relationship between the covariant anomaly and the bulk fractonic Hall response in \secref{SM:sec:fractonicAnomalyinflow}. 

\subsection{Derivation of edge current algebra}
Now, we derive the current algebra of the edge modes from the anomaly conditions \eqnref{SM:eqn:rank1anomaly} and \eqnref{SM:eqn:rank3anomaly}, following Ref. \cite{Wen_PRB1991}. It was argued in Ref. \cite{Wen_PRB1991}, in the context of fractional quantum Hall edge modes, that \eqnref{SM:eqn:rank1anomaly} for the current correlator, along with the assumption of locality implies that the edge modes must be gapless. This is because there is no ${K^{(1)}}^{\alpha \beta}\left(\omega,q\right)$ smooth near $\left(\omega, q\right)=0$ that satisfies \eqnref{SM:eqn:rank1anomaly}. Up to a polynomial, the current correlator ${K^{(1)}}^{\alpha \beta}\left(\omega,q\right)$ has the form
\begin{equation}\label{SM:eqn:K1form}
	{K^{(1)}}^{\alpha \beta}(\omega,q) = \frac{1}{2\pi k} \begin{cases}
		\frac{q}{\omega - v_1 q + i \delta}  &(\alpha, \beta) =(0,0) \\
		\half \frac{\omega + v_1 q }{\omega - v_1 q + i \delta}  & (\alpha, \beta) =(0,1) 
		, \text{or} \, (1,0) \\
		\frac{v_1 \omega}{\omega - v_1 q + i \delta} & (\alpha, \beta) =(1,1)
	\end{cases}
\end{equation}
The excitations have energies $\omega =v_1 q>0$, where $v_1$ is their velocity. The expectation value of the equal time commutator between the currents is
\begin{widetext}
\begin{equation}
	\begin{split}
		&\bra{0} [{j_q^{(1)}}^\alpha , j_{q'}^{(1)\beta}] \ket{0}= \lim_{t \to 0+} 
		\expect{{j_q^{(1)}}^\alpha\left(t\right) {j_{q'}^{(1)}}^\beta\left(0\right) - {j_{q'}^{(1)}}^\beta \left(t\right) {j_{q}^{(1)}}^\alpha\left(0\right) } \\
		&= \lim_{t \to 0^+} \frac{1}{L} \int \D{x_1} \D{x_1'} e^{-\ci q x_1 - \ci q' x_1'} \expect{{j^{(1)}}^ \alpha\left(t,x_1\right) {j^{(1)}}^ \beta\left(0,x_1'\right) - {j^{(1)}}^ \beta\left(t,x_1'\right) {j^{(1)}}^ \alpha\left(0,x_1\right)}\\
		&= \lim_{t \to 0^+} \frac{-\ci}{L} \int \D{x_1} \D{x_1'} e^{-\ci q x_1 - \ci q' x_1'} \left({K^{(1)}}^{\alpha \beta} \left(t,x_1-x_1'\right) - {K^{(1)}}^{\alpha \beta} \left(-t,x_1-x_1'\right) \right)\\
		&= \frac{-\ci}{2 \pi}\delta_{q+q', 0} \lim_{t \to 0^+} \int \D{\omega} \left( {K^{(1)}}^{\alpha \beta}\left(\omega,q\right) e^{-\ci \omega t} -{K^{(1)}}^{\alpha \beta} \left(\omega,q\right) e^{\ci \omega t} \right)
	\end{split}
\end{equation}
\end{widetext}
Using \eqnref{SM:eqn:K1form}, the expectation values can be calculated to be
\begin{equation}
	\begin{split}
		\expect{\left[{j_q^{(1)}}^0 , {j_{q'}^{(1)}}^  0\right]} &= \frac{-\ci}{2 \pi} \frac{1}{2\pi k } \delta_{q+q', 0} \lim_{t \to 0^+} \int \D{\omega} \left( e^{-\ci \omega \tau} \frac{q}{\omega -v_1 q +\ci \delta} \right. \\  & \qquad \qquad \qquad \qquad \qquad  \left. - e^{\ci \omega \tau} \frac{q}{\omega -v_1 q +\ci \delta} \right) \\ 
		&=-\frac{q}{2\pi k} \delta_{q +q',0}
	\end{split}
\end{equation}
Similarly, it can be shown that
\begin{equation}
	\frac{1}{v_1}\expect{\left[{j_q^{(1)}}^0 , {j_{q'}^{(1)  }}^1\right]}= \frac{1}{v_1^2}  \expect{\left[{j_q^{(1)}}^ 1 , {j_{q'}^{(1)  }}^1\right]} = -\frac{q}{2\pi k} \delta_{q +q',0}
\end{equation}
Defining $j^{(1) \pm} = \half \left( {j^{(1)}}^0 \pm \frac{1}{v_1} {j^{(1)}}^  1 \right)$, the commutators take the form
\begin{equation}
	\begin{split}\label{SM:eqn:KacMoodyRank1}
		\expect{\left[j_q^{(1)  +} , j_{q'}^{(1)  +}\right]} &=  -\frac{q}{2\pi k} \delta_{q +q',0} \\
		\expect{\left[j_q^{(1)  +} , j_{q'}^{(1)  -}\right]} &=  0\\ 
		\expect{\left[j_q^{(1)  -} , j_{q'}^{(1)  -}\right]} &=0
	\end{split}
\end{equation}
This is the well-known U$(1)$ Kac-Moody algebra \cite{Wen_PRL1990,Wen_PRB1991}, and was studied in the context of fractional quantum Hall edge physics. In our theory, this shows up as the  current algebra of non-fractonic transverse dipole modes on the edge. 

Now, let us focus on the rank$-3$ anomaly and derive the corresponding current algebra. There must be gapless modes associated with the rank$-3$ anomaly \eqnref{SM:eqn:rank3anomaly}. The current correlator ${K^{(3)}}^{\alpha \beta}\left(\omega , q\right)$ takes the form
\begin{equation}\label{SM:eqn:K3form}
	{K^{(3)}}^{\alpha \beta}\left(\omega,q\right) = \frac{1}{2\pi k} \begin{cases}
		\frac{q^3}{\omega - v_3 q^3 + i \delta}  &(\alpha, \beta) =(0,0) \\
		-\half \frac{\omega + v_3 q^3 }{\omega - v_3 q^3 + i \delta}  & (\alpha, \beta) =(0,1) 
		, \text{or} \, (1,0) \\
		\frac{v_3 \omega}{\omega - v_3 q^3 + i \delta} & (\alpha, \beta) =(1,1)
	\end{cases}
\end{equation}
In this case, the dispersion of the excitations is $\omega = v_3 q^3$. Note that \eqnref{SM:eqn:rank3anomaly} can not be satisfied if the excitations are linearly or quadratically dispersing. The minimum possible dynamical exponent for these edge modes is $z=3$ i.~e., cubically dispersing. Performing the calculation for the expectation values of the commutators of the rank$-3$ current operators, we find that
\begin{equation}
	\begin{split}
		\expect{\left[{j_q^{(3)}}^  0 , {j_{q'}^{(3)  }}^0\right]} &= -\frac{q^3}{2\pi k} \delta_{q+q',0} \\
		\expect{\left[{j_q^{(3)}}^  0 , {j_{q'}^{(3)}}^  1\right]} &= v_3 \frac{q^3}{2\pi k} \delta_{q+q',0} \\
		\expect{\left[{j_q^{(3)}}^ 1 , {j_{q'}^{(3)  }}^1\right]} &= -v_3^2\frac{q^3}{2\pi k} \delta_{q+q',0}
	\end{split}
\end{equation}
Defining $j^{(3) \pm} = \half \left( {j^{(3)}}^  0 \mp \frac{1}{v_3} {j^{(3)  }}^1 \right)$,
\begin{equation}
	\begin{split}\label{SM:eqn:KacMoodyRank3}
		\expect{\left[j_q^{(3)  +} , j_{q'}^{(3)  +}\right]} &=  -\frac{q^3}{2\pi k} \delta_{q +q',0} \\
		\expect{\left[j_q^{(3) +} , j_{q'}^{(3)  -}\right]} &=  0\\ 
		\expect{\left[j_q^{(3)  -} , j_{q'}^{(3) -}\right]} &=0
	\end{split}
\end{equation}
We call this new current algebra derived here as the fractonic U$(1)$ Kac-Moody algebra.

Although the results in \eqnref{SM:eqn:KacMoodyRank1} and \eqnref{SM:eqn:KacMoodyRank3} derived here concern the ground state expectation values of the commutators of the current operators, it was shown in Ref. \cite{Wen_PRB1991} that they also translate into operator identities under the assumption of locality.
\subsection{Edge theory: Representation of current algebra}
So far, we have obtained the commutation relations that must be satisfied by the current operators of the edge modes. In this section, we show how to write an action for the edge modes, so that the current operators of the resulting theory form a representation of the algebra Eqns.~\ref{SM:eqn:KacMoodyRank1}, \ref{SM:eqn:KacMoodyRank3}.

Edge theory corresponding to the algabra \eqnref{SM:eqn:KacMoodyRank1} is well-known, and can be written as
\begin{equation}\label{SM:eqn:edgetheoryrank1}
	S_{\textup{edge}}^{(1)}= \frac{k}{4\pi}\int \D{t} \, \D{x_1}\,  \left(\dou_t \psi^{(1)} \dou_1 \psi^{(1)} +v_1 \left(\dou_1 \psi^{(1)} \right)^2 \right)
\end{equation}
 Here, $\psi^{(1)}$ denotes a real bosonic field, and we require $k v_1<0$ for stability of the theory. The energy of the excitations is $\omega =v_1 q$. The conjugate momentum is $\Pi^{(1)}\left(x_1\right) = \frac{k}{4\pi} \dou_1 \psi^{(1)} \left(x_1\right)$. To see that the current operators of this theory indeed satisfy the required commutation relations, we quantize the theory to find that
 \begin{equation}
	\begin{split}
		[\psi^{(1)} \left(x_1\right), \psi^{(1)}\left(x_1'\right) ] &= -\ci \frac{\pi}{k} \text{sgn}\left(x_1-x_1'\right)\\
		[\dou_1 \psi^{(1)} \left(x_1\right), \dou_{1'}\psi^{(1)}\left(x_1'\right) ] &= \ci \frac{2\pi}{k} \dou_1 \delta\left(x_1-x_1'\right)		
	\end{split}
\end{equation}
Now, with ${j^{(1)}}^0 = \frac{1}{v_1} {j^{(1)}}^1 = \frac{1}{2\pi} \dou_1 \psi^{(1)}$,
\begin{equation}
	[{j^{(1)}}^0\left(x_1\right), {j^{(1)}}^0\left(x_1'\right)] = \ci \frac{1}{2\pi k} \dou_1\delta\left(x_1 - x_1'\right)
\end{equation}
The above commutation relation is equivalent to the current algebra \eqnref{SM:eqn:KacMoodyRank1} in Fourier space. Let us now consider the operator $e^{\ci D^\perp \psi^{(1)}\left(x_1'\right)}$ and its commutation relation with the density operator:
\begin{equation}
[{j^{(1)}}^0\left(x_1\right), e^{\ci D^\perp \psi^{(1)}\left(x_1'\right)}] = \frac{D^\perp}{k} \delta\left(x_1-x_1'\right) e^{\ci D^\perp \psi^{(1)}\left(x_1'\right)}
\end{equation}
This confirms that the operator $\exp{\ci D^\perp \psi^{(1)}\left(x'\right)}$ creates a transverse dipole of strength $\frac{D^\perp}{k}$ (as coupled to background gauge fields) at the location $x'$ on the edge.

Now, let us turn to the fractonic current algebra \eqnref{SM:eqn:KacMoodyRank3}. These edge modes couple to rank$-3$ gauge fields, and the \eqnref{SM:eqn:consistentAnomalyContinuity} suggests that the charge, dipole moment, and quadrupole moment are conserved in the absence of background electric fields. With this motivation, we consider a real bosonic field $\psi^{(3)} \left(x_1\right)$ and we would like to write an action of this bosonic field that is invariant under the transformation $\psi^{(3)} \to \psi^{(3)} + a + bx_1 + cx^2_1$ for arbitrary real constants $a,b,c$. The simplest quantity invariant under this transformation is $\dou^3_1 \psi^{(3)}$. Based on this discussion, we write the action
\begin{equation}
	S_{\textup{edge}}^{(3)}= \frac{k}{4\pi}\int \D{t} \D{x_1} \left(-\dou_t \psi^{(3)} \dou_1^3 \psi^{(3)} +v_3 \left(\dou_1^3 \psi^{(3)} \right)^2 \right).
\end{equation}
Again, we require $k v_3<0$ for stability.
The conjugate momentum is $\Pi^{(3)} \left(x_1\right) = -\frac{k}{4\pi} \dou_1^3 \psi^{(3)} \left(x_1\right)$. Quantizing the theory using the method of Dirac brackets leads to the commutator
\begin{equation}
	[\psi^{(3)} \left(x_1\right), \psi^{(3)}\left(x_1'\right) ] = \ci \frac{\pi}{2k} \left(x_1-x_1'\right)^2 \text{sgn}\left(x_1-x_1'\right)
\end{equation}
Also, note the following commutation relations:
\begin{equation}
	\begin{split}
		[\dou_1 \psi^{(3)}\left(x_1\right), \dou_{1'}\psi^{(3)}\left(x_1'\right)] &= -\ci \frac{\pi}{k} \textup{sgn}\left(x_1-x_1'\right) \\
		[\dou_1^2 \psi^{(3)}\left(x_1\right), \dou_{1'}^2 \psi^{(3)}\left(x_1'\right)] &=  \ci \frac{2\pi}{k} \dou_1 \delta(x_1-x_1') \\
		[\dou_1^3 \psi^{(3)}\left(x_1\right), \dou_{1'}^3 \psi^{(3)}\left(x_1'\right)] &=  -\ci \frac{2\pi}{k} \dou_1^3 \delta(x_1-x_1')
	\end{split}
\end{equation}
Thus, with ${j^{(3)}}^0 = -\frac{1}{v_3} {j^{(3)}}^1 = \frac{1}{2\pi} \dou_1^3 \psi^{(3)}$, we see that \eqnref{SM:eqn:KacMoodyRank3} is satisfied. Thus, ${j^{(3)}}^0 \left(x_1\right)$ is the local fracton charge density on the edge.

Finally, we identify the vertex operators that create point excitations on the edge. The vertex  operators $e^{\ci Q v \psi^{(3)}\left(x_1'\right)}$, $e^{\ci D \dou_{1'}\psi^{(3)}\left(x_1'\right)}$ and $e^{\ci  \Theta \ell \dou_{1'}^2 \psi^{(3)}\left(x_1'\right)}$ have the following commutation relations with the density operator ${j^{(3)}}^0\left(x\right)$.
\begin{equation}
	\begin{split}
		[{j^{(3)}}^0\left(x\right), e^{\ci  Q \ell^{-1} \psi^{(3)}\left(x_1'\right)}] &= - \ell^{-1}\frac{ Q}{k} \delta\left(x_1-x_1'\right) e^{\ci  Q \ell^{-1} \psi^{(3)}\left(x_1'\right)} \\
		[{j^{(3)}}^0\left(x\right), e^{\ci D \dou_{1'}\psi^{(3)}\left(x_1'\right)}] &= \frac{D}{k} \dou_1 \delta\left(x_1-x_1'\right)  e^{\ci D \dou_{1'}\psi^{(3)}\left(x_1'\right)} \\
		[{j^{(3)}}^0\left(x\right), e^{\ci  \Theta \ell \dou_{1'}^2 \psi^{(3)}\left(x_1'\right)}] &= \ell \frac{ \Theta}{k} \dou_1^2 \delta\left(x_1-x_1'\right) e^{\ci  \Theta \ell \dou_{1'}^2 \psi^{(3)}\left(x_1'\right)}
	\end{split}
\end{equation}
 Hence, these operators create point charges of strength $-\ell ^{-1}\frac{Q}{k}$ , point longitudinal dipoles of strength $-\frac{D}{k}$ and point quadrupoles of strength $\ell \frac{ \Theta}{k}$ respectively at the location $x'$ on the edge.
 
 \section{Fractonic Hall response and the edge anomaly}\label{SM:sec:fractonicAnomalyinflow}

 In fractional quantum Hall systems, it is known that \cite{Stone1991} the bulk Hall response compensates for the non-conservation of charge on the edge. In this section, we explore this relation in the context of FCS theory, and show how the anomalies of the edge degrees of freedom (\eqnref{SM:eqn:covariantAnomalyContinuity}) conspire to produce the fractonic Hall response(\eqnref{eqn:hallresponse}, main text). 
 
We consider the system to be an infinite bar, along the $x_1$ direction, as shown in \figref{fig:anomaly_transport}, main text. The covariant anomaly for the top edge is given by \eqnref{SM:eqn:covariantAnomalyContinuity}, while for the bottom edge, it is of the same form with $k$ replaced by $-k$ (opposite chirality). In terms of the background  fields $E_{11},E_{12}$, the rates of change of the total charge ${\cal Q}_\pm$, longitudinal dipole moment ${\cal D}_\pm$ obtained from the anomaly of fractonic edge modes is
\begin{equation}\label{SM:eqn:anomalyinflow:F}
	\begin{split}
    \frac{\D{{\cal Q}_\pm}}{\D{t}} &=\pm \frac{1}{2\pi k} \int \D{x_1} \dou_2 E_{12}  \\
    \frac{\D{{\cal D}_\pm}}{\D{t}} &= \mp\frac{1}{\pi k} \int \D{x_1} E_{11} \pm \frac{1}{2\pi k} \int \D{x_1} x_1 \dou_2 E_{12}. 
	\end{split}
\end{equation}
Similarly, the rate of change transverse dipole moment ${\cal D}^\perp_\pm$ arising from the anomaly of the non-fractonic edge modes is 
\begin{equation}\label{SM:eqn:anomalyinflow:NF}
	\begin{split}
    \frac{\D{{\cal D}^\perp_\pm}}{\D{t}} &= \mp \frac{1}{2\pi k} \int \D{x_1} E_{12}.
	\end{split}
\end{equation}
In the above $\pm$ stands for top and bottom edges of \figref{fig:anomaly_transport}.
We discuss various choices of background electric fields to uncover the connection between the fractonic Hall response, and the above edge anomalies \eqnref{SM:eqn:anomalyinflow:F} and \eqnref{SM:eqn:anomalyinflow:NF}.

\paraheading{Response to $E_{11}$}
We consider the background electric field tensor $E_{ij}$ with $E_{12}=0$ and spatially uniform $ E_{11}=-E_{22}$. From \eqnref{SM:eqn:anomalyinflow:F}, we see that the density of longitudinal dipoles on the top (bottom) edge changes at the rate $\mp\frac{E_{11}}{\pi k}$. This must be compensated by a flux $\pm \Sigma$ of longitudinal dipoles  on the top(bottom) edges, where $\Sigma = -\frac{E_{11}}{\pi k}$. This flux is realized by a current of longitudinal dipoles (see \figref{fig:anomaly_transport}) given by 
\beq
J_{ij} \equiv \half \begin{pmatrix} 0 & \Sigma \\
\Sigma &0 
\end{pmatrix}
\eeq
This current of dipoles is not only consistent with the fractonic constraints, but also with the predicted bulk response.

This illustrates how the anomaly of the fractonic edge mode (longitudinal dipoles) is connected to the bulk Hall response. 

\paraheading{Response to $E_{12}$} Now, we consider a uniform electric field  $E_{11}=E_{22}=0$, $E_{12} \ne 0$. 

According to the edge anomaly \eqnref{SM:eqn:anomalyinflow:NF}, the rate of change of transverse dipole density on the top (bottom) edges is $\mp \frac{1}{2 \pi k} E_{12}$. This can be accomplished by a flux $\Sigma = - \frac{1}{2 \pi k} E_{12}$ of transverse dipoles on the top edge and $-\Sigma$ on the bottom edge. Although transverse dipoles cannot move normal to the edge, a bulk current that produces such a $\Sigma$ consistent with fractonic constraints can be realized by dipoles oriented at $\pm 45^\circ$ as shown in \figref{fig:anomaly_transport}. In the frame $\tilde{x}_1-\tilde{x}_2$ (\figref{fig:anomaly_transport}), this current is 
\beq
\tilde{J}_{ij} \equiv  \begin{pmatrix}
 0 & -\Sigma \\
 -\Sigma & 0
\end{pmatrix}
\eeq
which when transformed to the $x_1-x_2$ frame gives
\beq
J_{ij} \equiv \begin{pmatrix}
 -\Sigma & 0 \\
 0 & \Sigma 
\end{pmatrix}
\eeq
leading to $J_{11} = \frac{1}{2 \pi k} E_{12}$, which is precisely the fractonic Hall response. We have thus established the relationship between the anomaly of the non-fractonic modes and the bulk Hall response.

\section{Edge-to-edge tunneling in the fracton Hall bar}\label{SM:sec:StatbilityRG}

In this section, we study local edge-to-edge tunneling \cite{MoonKaneGirvinFischer1993} between the edges of a fracton Hall bar. We give the details concerning the renormalization group analysis of this problem.

Consider an infinite fracton bar (\figref{fig:edge_stability}) described by the fracton Chern-Simons theory. Note that charges can not tunnel from one edge to the other because of the conservation of the dipole moment. Interestingly, the tunneling of longitudinal dipoles is the simplest process obeying the fractonic constraints. We write a lagrangian capturing these allowed processes. Let the tunneling between the edges be allowed to occur at the point $x_1=0$, and denote the bosonic fields of the top (+) and the bottom (-) edges as $\psi^{(3)}_\pm$ respectively.
\begin{equation}
	\begin{split}
		S= \frac{k}{4\pi} &\int \D{t} \D{x_1}   \sum_\sigma \left( -\sigma \dou_t \psi^{(3)}_{\sigma} \dou_1^3 \psi^{(3)}_{\sigma}   + v_3 \left(\dou_1^3 \psi^{(3)}_{\sigma} \right)^2  \right) \\  - &\sum_{D} g_D \int \D{t} \cos\left[{D\left(\dou_1 \psi^{(3)}_+\left(t,0\right) - \dou_1 \psi^{(3)}_-\left(t,0\right)\right) }\right]
	\end{split}
\end{equation}
Making a change of variables by
\begin{equation}
	\begin{split}
		\psi_+ - \psi_- &= \psi \\
		\frac{\psi_+ + \psi_-}{2} &= \tilde{\psi},
	\end{split}
\end{equation}
and integrating out $\tilde{\psi}$ fields allows us to write the action in euclidean time as
\begin{equation}
	S=-\frac{k}{8\pi} \int \D{\tau} \D{x_1} \frac{\left(\dou_t \psi \right)^2}{v^{(3)}} + v_3 \left(\dou_1^3 \psi\right)^2 + \sum_D g_D \int \D{\tau} \cos{D \dou_1 \psi(\tau,0)}.
\end{equation}
As usual, we have $kv_3<0$ for the stability of the theory. 
At the fixed point given by $g_m=0$, the scale transformation is given by $x'=x/\zeta$, $\tau'=\tau/\zeta^3$. Let us perform Wilsonian renormalization group analysis about this fixed point. At a given momentum cutoff $\Lambda$, let the modes in the frequency and momentum space be given by $\frac{\omega^2}{v_3^2} + q^6 <\Lambda^6$. We want to integrate out the modes $\psi_>$ in the region $R(\Lambda, \zeta)$ satisfying $\left(\frac{\Lambda}{\zeta}\right)^6 <\frac{\omega^2}{v_3^2}  + q^6 < \Lambda^6$, to find an action of the long wavelength modes $\psi_<$.
We want 
\begin{equation}
	\begin{split}
		\expect{S_I}_> &= \sum_D g_D \int \D{\tau} \expect{\cos{D \dou_x \psi }}_> \\
		&= \sum_D g_D \int  \D{\tau} e^{-\frac{D^2}{2} \expect{\left(\dou_x \psi_> \right)^2}_>} \cos{D \dou_x \psi_<}
	\end{split}
\end{equation}
\begin{widetext}
We have
\begin{equation}
	\expect{\psi_> (\tau,x_1) \psi_> (\tau',x_1')}_>=- \frac{1}{\left(2\pi\right)^2 } \frac{4\pi}{k} \int_{R\left(\Lambda, \zeta\right)}  dq d \omega \frac{e^{i q \left(x-x'\right) + i \omega \left(\tau - \tau'\right)}}{\frac{\omega^2}{v_3} + v_3 q^6}
\end{equation}
\begin{equation}
	\expect{\dou_1 \psi_>(\tau,x_1) \dou_{1'} \psi_>(\tau',x_1')}_> = -\frac{1}{\left(2\pi\right)^2 } \frac{4\pi}{k} \int_{R\left(\Lambda, \zeta\right)}  dq d \omega \frac{ q^2 e^{i q \left(x-x'\right) + i \omega \left(\tau - \tau'\right)}}{\frac{\omega^2}{v_3} + v_3 q^6}
\end{equation}
\end{widetext}
Changing to the dimensionless variables $\upsilon=\frac{q^3}{\Lambda^3}$, $\chi=\frac{\omega}{v_3 \Lambda^3}$,
\begin{equation}
	\begin{split}
\expect{\left(\dou_1 \psi_>\right)^2}_> &= \frac{1}{\left(2\pi\right)^2 } \frac{4\pi}{\abs{k}} \times \frac{1}{3} \int_{1/\zeta^6 < \upsilon^2+\chi^2 < 1} \D{\upsilon} \D{\chi} \frac{1}{\upsilon^2 +\chi^2} \\ 
		&= \frac{1}{3\pi \abs{k} } \times 2\pi \log{\zeta^3} = \frac{2}{\abs{k}} \log{\zeta}
	\end{split}
\end{equation}
Now, 
\begin{equation}
	\expect{S_I}_> = \sum_D g_D \int \D{\tau} e^{-\frac{D^2}{\abs{k}} \log{\zeta}} \cos{D \dou_1 \psi_<}
\end{equation}
Finally rescaling the cutoff, or $\tau' = \tau/\zeta^3$, $x'=x/\zeta$, and writing $\zeta=e^l$,
\begin{equation}
	\frac{\text{d} g_D}{\text{d} l} = \left(3-\frac{D^2}{\abs{k}} \right) g_D.
\end{equation}

\fi 

\end{document}